\documentclass{article}

\usepackage{graphicx}
\usepackage{epsfig}
\usepackage{authblk}
\usepackage{subfigure}
\usepackage{wrapfig}
\usepackage[T1]{fontenc}
\usepackage[latin9]{inputenc}
\usepackage{units}
\usepackage{amsmath}
\usepackage{amssymb}
\usepackage{esint}
\usepackage{babel}
\usepackage{color}
\usepackage{float}
\usepackage{lipsum}  
\usepackage{fancyhdr}
\usepackage{lineno}

\usepackage{tikz}

\title{Triple Hill's Vortex Synthetic Eddy Method
}

\author[1]{John S. Haywood \thanks{email: jsh478@cavs.msstate.edu}}
\author[1]{Adrian Sescu \thanks{email: sescu@ae.msstate.edu}}
\author[1]{Shanti Bhushan}
\author[2]{Chris Kees}

\affil[1]{Mississippi State University, MS 39762, USA}
\affil[2]{US Army Engineer Research and Development Center (ERDC), Vicksburg, MS}

\begin{document}

\date{}

\maketitle

\begin{abstract}
The generation of initial or inflow synthetic turbulent velocity or scalar fields reproducing statistical characteristics of realistic turbulence is still a challenge. The synthetic eddy method, previously introduced in the context of inflow conditions for large eddy simulations, is based on the assumption that turbulence can be regarded as a superposition of coherent structures. In this paper, a new type of synthetic eddy method is proposed, where the fundamental eddy is constructed by superposing three Hill's vortices, with their axes orthogonal to each other. A distribution of Hill's vortices is used to synthesize an anisotropic turbulent velocity field that satisfies the incompressibility condition and match a given Reynolds stress tensor.  The amplitudes of the three vortices that form the fundamental eddy are calculated from known Reynolds stress profiles through a transformation from the physical reference frame to the principal-axis reference frame. In this way, divergence-free anisotropic turbulent velocity fields are obtained that can reproduce a given Reynolds stress tensor. The model was tested on both isotropic and anisotropic turbulent velocity fields, in the framework of grid turbulence decay and turbulent channel flow, respectively. The transition from artificial to realistic turbulence in the proximity to the inflow boundary was found to be small in all test cases that were considered.
  
\end{abstract}

\section{Introduction}
There is a wide range of models for generating turbulent velocity fields at the LES inflow boundary, roughly divided into several classes \cite{Shur}: precursor DNS or LES; turbulence recycling; synthetic turbulence; artificial forcing or volume source terms; and vortex generating techniques. The most accurate technique to impose turbulent inflow conditions is based on a precursor simulation performed additionally in a separate flow domain with periodic boundary conditions in the streamwise direction; data is extracted from a cross-sectional plane and imposed at the inflow boundary of the main simulation. This method is known to produce realistic turbulent structures at the inflow boundary and is able to capture the correct flow physics in the flow region of interest. However, this technique has two major drawbacks: firstly, it is restricted to simple applications where the flow at the inflow boundary can be regarded as a fully developed turbulent flow \cite{Kaltenbach} or a spatially developed turbulent boundary layer \cite{Lund}. Secondly, this technique is computationally expensive because two simulations have to be carried out. 

Another choice is to employ recycling techniques by which the streamwise periodic conditions of developing flows are modified to account for the streamwise variations of the flow \cite{Lund}. Spalart \cite{Spalart} developed such a method to account for spatial growth in boundary layers, by using periodic boundary conditions in the streamwise direction and source terms in the Navier-Stokes equations to account for the growth. Spalart's method is ideal to generate inflow boundary conditions because it is capable of producing a spatially evolving boundary layer. Methods based on proper orthogonal decomposition \cite{Druault,Johansson} are less expensive than the other methods, but they require appropriate DNS \cite{Johansson} or high-fidelity experimental data \cite{Druault} to obtain the energetic modes.

Synthetic turbulence modeling is an attractive and inexpensive approach that is utilized to generate initial or inflow conditions for Large Eddy Simulations (LES) or Direct Numerical Simulations (DNS), or to analyze fundamental structures of turbulent flows. To be feasible, a synthetic turbulence approach must ensure that specific statistics from the target turbulent data are reproduced. A number of proposed synthetic turbulence models \cite{Li,Bechara,Fung} are based on spectral methods, which are mostly utilized to generate isotropic turbulent flow-fields (the pioneering work of Kraichnan \cite{Kraichnan} is worth mentioning here). The resulting turbulent kinetic energy in this case is uniformly spread over all wavenumbers, and as a result the pseudo-turbulence is rapidly dissipated, and a long distance is required to recover the realistic turbulent flow. The model of Maxey \cite{Maxey} is a successful application of Kraichnan's method to applications involving anisotropic turbulent flows. His technique is based on filtering and scaling operations applied to the flow field. Smirnov et al. \cite{Smirnov} extended the method proposed by Maxey \cite{Maxey} by using scaling and simple coordinate transformation operations, which were found to be efficient and fast. Based on the work of Smirnov et al. \cite{Smirnov}, Batten et al. \cite{Batten} reconstructed an inflow turbulent fluctuation flow field by summing a set of Fourier modes with prescribed random phases and amplitudes, and preserving the space/time correlations and the second order moments. Keating et al. \cite{Keating} used Batten's model to generate inflow boundary conditions for a plane channel flow and found that the development of the turbulent flow was slow: after a distance of ten channel heights the turbulent flow in the core region was clearly not yet fully developed. Billson et al. \cite{Billson} used the idea of scaling and coordinate transformation as proposed by Smirnov et al. \cite{Smirnov} to generate anisotropic synthesized velocity fields from isotropic Reynolds stress tensors. Davidson and Billson \cite{Davidson} then applied this method to fully-developed channel flow. Rosales and Meneveau \cite{Rosales_1,Rosales_2} proposed a simple method, called multiscale minimal Lagrangian map (MMLM), to generate synthetic vector fields, starting with an initial Gaussian field and using deformed random-phase Fourier modes. Klein et al. \cite{Klein} developed a synthetic turbulence model that reproduced first and second order one-point statistics and local correlations, based on three-dimensional digital filters. The temporal correlations were well-predicted by shifting the random data in time. Kempf et al. \cite{Kempf_1} showed that a method solving for an appropriate diffusion equation using three-dimensional random data was equivalent to applying a filter based on Gaussian functions. Using this idea, a new synthetic turbulence model was devised which can be easily implemented in both structured and unstructured grid codes. These previous methods approach the modeling of realistic turbulence from a statistical point of view, modify fields of random fluctuations as a means to match as many turbulent statistics as possible. There is another class of method that is based in a physical description of turbulence and tries to model the coherent structures present in turbulence through the combination of discrete synthetic structures.

A synthetic eddy method (SEM) based on the assumption that turbulence can be considered as a superposition of coherent structures was proposed by Jarrin et al. \cite{Jarrin_1}. The eddies were convected downstream with the mean flow, through the inlet plane. Jarrin's synthetic eddy method showed similar behavior as the other existent models based on Fourier modes or digital filters, especially for spatial decay of isotropic turbulence and turbulent channel flows. Pami\`es et al. \cite{Pamies} finely tuned the shape functions to mimic the elongated vortices observed in the near-wall region. Roidl et al. \cite{Roidl} then tuned the shape functions in the logarithmic and wake regions for a turbulent boundary layer. Marked improvements in the reduction of the development length were seen as more realistic modeling of the coherent structures was included in the imposed synthetic eddies. In order to further reduce the development length, Poletto et al. \cite{Poletto_1} applied the synthetic eddy framework to the vorticity field and solved for the fluctuating velocity components using the Biot-Savart law. From this solution process, restrictions on the shape functions were found that were then used to select a function that satisfied the divergence-free condition. This formulation only includes one shape function, based on a sine function, and one characteristic eddy length scale, and as such, could only match a limited range of anisotropic Reynolds stress tensors. To be able to reproduce a wide range of anisotropic turbulence, Poletto et al. \cite{Poletto_2} reformulated the method to allow for separate shape functions and lengths scales for each dimension. This destroyed the satisfaction of the divergence-free condition guaranteed by the previous formulation, but by taking the divergence of the prospective eddy velocity field sufficient conditions were found to be able to select divergence-free shape functions, based on a quadratic polynomial. Poletto et al. \cite{Poletto_2} showed that pressure fluctuations at the inlet, caused by the divergence-free condition not being satisfied, were greatly reduced when compared with the original synthetic eddy method for plane channel flow. Skillen et al. \cite{Skillen} introduced shape functions that are scaled with the local synthetic eddy population density and also modified the homogeneous eddy location distribution of the original SEM to allow for clustering of smaller synthetic eddies near the wall. In the context of computational aeroacoustics, a divergence-free synthetic eddy model was constructed by Sescu and Hixon \cite{Sescu} based on the vector potential instead of the velocity field.

In this work, the proposed Triple Hill's Vortex (THV) synthetic eddy method builds upon the foundation of the current SEM's, but instead of using abstract mathematical functions to determine the shape of eddies, the THV SEM uses a superposition of Hill's spherical vortices. Since the Hill's spherical vortex is an actual solution to the Euler equation, the synthetic inflow velocity field will reproduce more realistic turbulent flow conditions. Specifically, within the proposed THV model, three orthogonal Hill's vortices are superposed, with the amplitudes of the individual vortices determined by using a transformation from the physical reference frame to the principal-axis frame, as outlined in Billson et al. \cite{Billson}.

\section{Governing Equations}
The governing equations employed consist of the Favre-filtered full compressible Navier-Stokes equations written in curvilinear coordinates and conservative form. A generalized curvilinear coordinate transformation in the three-dimensional form 
$\xi = \xi \left(x,y,z \right),
\eta = \eta \left(x,y,z \right),
\zeta = \zeta \left(x,y,z \right)$,
is considered, where $\xi$, $\eta$, and $\zeta$ are the spatial coordinates in the computational space, and $x$, $y$, and $z$ are the spatial coordinates in physical space. As a note, a tilde ($\widetilde{\quad}$) represents Favre-filtering at the grid level $\Delta$. In conservative form, the Navier-Stokes equations are written as

\begin{eqnarray}\label{NS}
\mathbf{Q}_t
+ \mathbf{F} _{\xi}
+ \mathbf{G}_{\eta}
+ \mathbf{H}_{\zeta}
= \mathbf{S}
\end{eqnarray}
where the vector of conservative variables is given by

\begin{equation}
\mathbf{Q} = \frac{1}{J} \{ 
\begin{array}{rrrrrr}
\overline{\rho},   \hspace{4mm}
\overline{\rho}\widetilde{u}_i,   \hspace{4mm}
\overline{\rho}\widetilde{E}
\end{array}
\}^{T},\quad i = 1,2,3
\end{equation}
$\overline{\rho}$ is the mean density of the fluid, $\widetilde{u}_i = (\widetilde{u}, \widetilde{v}, \widetilde{w})$ is the filtered velocity vector in physical space, and $\overline{\rho}\widetilde{E}$ is the total energy. The flux vectors, $\mathbf{F}$, $\mathbf{G}$, and $\mathbf{H}$, are given by

\begin{eqnarray}
\mathbf{F} = \frac{1}{J} \left\{ 
\begin{array}{c}
\overline{\rho} U,   \hspace{4mm}
\overline{\rho} \widetilde{u}_i U + \xi_{x_i} (\overline{p} + \tau_{i1}),   \hspace{4mm}
\overline{\rho}\widetilde{E} U + \overline{p} U +  \xi_{x_i} \Theta_i
\end{array}
\right\}^{T}, \\
\mathbf{G} = \frac{1}{J} \left\{ 
\begin{array}{c}
\overline{\rho} V,   \hspace{4mm}
\overline{\rho} \widetilde{u}_i V + \eta_{x_i} (\overline{p} + \tau_{i2}),    \hspace{4mm}
\overline{\rho}\widetilde{E} V + \overline{p} V+  \eta_{x_i} \Theta_i
\end{array}
\right\}^{T},  \\ 
\mathbf{H} = \frac{1}{J} \left\{ 
\begin{array}{c}
\overline{\rho} W,   \hspace{4mm}
\overline{\rho} \widetilde{u}_iW + \zeta_{x_i} (\overline{p} + \tau_{i3}),    \hspace{4mm}
\overline{\rho}\widetilde{E} W + \overline{p} W +  \zeta_{x_i} \Theta_i
\end{array}
\right\}^{T}
\end{eqnarray}
where the contravariant velocity components are given by

\begin{eqnarray}
U = \xi_{x_i} \widetilde{u}_i ,   \hspace{4mm}
V = \eta_{x_i} \widetilde{u}_i,    \hspace{4mm}
W = \zeta_{x_i} \widetilde{u}_i
\end{eqnarray}
with the Einstein summation convention applied over $i = 1,2,3$, the shear stress tensor and the heat flux are given as

\begin{equation}
\tau_{ij} = \frac{\widetilde{\mu}}{Re} \left[
\left(
\frac{\partial \xi_k}{\partial x_j}  \frac{\partial \widetilde{u}_i}{\partial \xi_k}  +
\frac{\partial \xi_k}{\partial x_i}  \frac{\partial \widetilde{u}_j}{\partial \xi_k}
\right)
- \frac{2}{3} \delta_{ij} \frac{\partial \xi_l}{\partial x_k}  \frac{\partial \widetilde{u}_k}{\partial \xi_l}
\right] + \tau_{ij}^{sgs}
\end{equation}

\begin{equation}
\Theta_{i} = 
 \widetilde{u}_j \tau_{ij} + \frac{\widetilde{\mu}}{(\gamma-1)M_{\infty}^2 Re Pr}
\frac{\partial \xi_l}{\partial x_i}  \frac{\partial \widetilde{T}}{\partial \xi_l} + q_i^{sgs}
\end{equation}
respectively. $\tau_{ij}^{sgs}$ and $q_{i}^{sgs}$ are the subgrid scale stress and subgrid scale heat flux terms, which are defined in Section \ref{section_SGS}. The pressure $\overline{p}$, the temperature $\widetilde{T}$, and the density of the fluid are combined in the equation of state, $\overline{p} = \overline{\rho} \widetilde{T} / \gamma M_{\infty}^2$. Other notations include the dynamic viscosity $\mu$, Reynold's number $Re=\rho_{\infty} V_{\infty}L/\mu$ based on a characteristic velocity $V_{\infty}$, and a characteristic length $L$, the free-stream Mach number $M_{\infty}=V_{\infty}/a$ (with $a$ being the speed of sound), Prandtl's number $Pr = C_p \mu/k$ (where $k$ is thermal conductivity), the specific heat at constant pressure $C_p$,  and the ratio between the specific heats $\gamma$. The Jacobian of the curvilinear transformation from the physical space to computational space is denoted by $J$. The derivatives $\xi_x$, $\xi_y$, $\xi_z$, $\eta_x$, $\eta_y$, $\eta_z$, $\zeta_x$, $\zeta_y$, and $\zeta_z$ represent grid metrics. The variables are non-dimensionalized by their respective free-stream variables, except from pressure which is non-dimensionalized by $\rho_{\infty} V_{\infty}$. The dynamic viscosity and thermal conductivity are linked to the temperature using the Sutherland's equations in dimensionless form,

\begin{eqnarray}
\widetilde{\mu} = \widetilde{T}^{3/2} \frac{1 + C_1/T_{\infty}}{\widetilde{T}+C_1/T_{\infty}}; \hspace{4mm}
\widetilde{k} = \widetilde{T}^{3/2} \frac{1 + C_2/T_{\infty}}{\widetilde{T}+C_2/T_{\infty}},
\end{eqnarray}
where for air at sea level, $C_1 = 110.4 K$, $C_2 = 194 K$, and $T_{\infty}$ is a reference temperature.

\subsection{Subgrid Scale Model}\label{section_SGS}
The SGS stress is modeled using the Coherent Structure Model (CSM) developed by Kobayashi \cite{Kobayashi_1} for incompressible flow and later applied to compressible flows by Hadjadj et al. \cite{Hadjadj} and Ben-Nasr et al. \cite{Ben_Nasr}. The CSM is based on the assumption that the SGS dissipation is small at the center of a coherent eddy and that the energy transfer between the resolved scales and the SGS occurs around the edge of this coherent eddy \cite{Kobayashi_1,Kobayashi_2}.

The SGS stress tensor, $\tau_{ij}^{sgs} = \overline{\rho}\left( \widetilde{u_i u_j} - \widetilde{u_i}\widetilde{u_j} \right)$, is defined using an eddy-viscosity model as as follows
\begin{equation}
\tau_{ij}^{sgs} = -2 \mu_{sgs} \left( \widetilde{S}_{ij} - \frac{1}{3}\widetilde{S}_{kk} \delta_{ij} \right) - \frac{1}{3}\tau_{kk}^{sgs}\delta_{ij}
\end{equation}
where $\widetilde{S}_{ij}$ is the resolved strain-rate tensor. The SGS viscosity, $\mu_{sgs}$, is defined as
\begin{equation}
\mu_{sgs} = \overline{\rho} C_s \Delta^2 |\widetilde{S}|
\end{equation}
where $\Delta$ is the grid scale, $C_s$ is the dynamically calculated Smagorinsky coefficient, and $|\widetilde{S}|$ is the strain-rate magnitude. The isotropic part of the SGS tensor, $\tau_{kk}^{sgs}$, is modeled using the following relationship proposed by Yoshisawa \cite{Yoshisawa}
\begin{equation}
\tau_{kk}^{sgs} = 2 \overline{\rho} C_I \Delta^2 |\widetilde{S}|^2
\end{equation}
where $C_I$ is another dynamically calculated model coefficient.

For the CSM, the Smagorinsky coefficient is defined as
\begin{equation}
C_s = C_{csm} |F_{cs}|^{3/2} \left( 1 - F_{cs} \right)
\end{equation}
where $C_{csm}$ is the CSM model coefficient, equal to $1/22$ \cite{Kobayashi_1,Kobayashi_2}, and $F_{cs}$ is the coherent structure function.
\begin{equation}
F_{cs} = \frac{\widetilde{Q}}{\widetilde{E}}
\end{equation}
$\widetilde{Q}$ is the second invariant of the resolved velocity gradient and $\widetilde{E}$ is the magnitude of the resolved velocity gradient tensor.
\begin{eqnarray}
\widetilde{Q} &= \frac{1}{2} \left( \widetilde{W}_{ij}\widetilde{W}_{ij} - \widetilde{S}_{ij}\widetilde{S}_{ij}\right) =& -\frac{1}{2} \frac{\partial \widetilde{u}_j}{\partial x_i}\frac{\partial \widetilde{u}_i}{\partial x_j} \\
\widetilde{E} &= \frac{1}{2} \left( \widetilde{W}_{ij}\widetilde{W}_{ij} + \widetilde{S}_{ij}\widetilde{S}_{ij}\right) =& \ \ \ \frac{1}{2} \frac{\partial \widetilde{u}_j}{\partial x_i}\frac{\partial \widetilde{u}_j}{\partial x_i}
\end{eqnarray}
$\widetilde{S}_{ij}$ is the resolved strain-rate tensor and $\widetilde{W}_{ij}$ is the resolved vorticity tensor.

The coefficient for the isotropic part of the SGS tensor, $C_I$, is dynamically modeled in a similar fashion to the Smagorinsky coefficient.
\begin{equation}
C_I = C_{csm_I} |F_{cs}|^{3/2} \left( 1 - F_{cs} \right)
\end{equation}
$C_{csm_I}$ is the CSM isotropic model coefficient and is chosen here such that the maximum value of $C_I$ is $0.0066$ \cite{Zang}

The SGS heat flux term, $q_{i}^{sgs}$, is also modeled using an eddy-viscosity model.
\begin{equation}
q_i^{sgs} = - \frac{\mu_{sgs}}{\left(\gamma - 1 \right) M_{\infty}^2 Re Pr_{sgs}} \frac{\partial \widetilde{T}}{\partial x_i}
\end{equation}
The SGS Prandtl number, $Pr_{sgs}$, is given the constant value of $0.7$ \cite{Zang,Martin}.

\section{Numerical Algorithm}
The compressible Navier-Stokes equations are solved in the framework of Large Eddy Simulations, where the Coherent Structure Model is applied to account for the missing sub-grid scale energy. The numerical algorithm uses high-order finite difference approximations for the spatial derivatives and explicit time marching. The time integration is performed using a fully-explicit fourth-order Adams-Bashforth scheme \cite{Butcher} of the form
\begin{eqnarray}\label{28}
\mathbf{Q}^{n+1} = \mathbf{Q}^{n} + \Delta t \left[ \frac{55}{24}L\left( \mathbf{Q}^{n} \right) - \frac{59}{24}L\left( \mathbf{Q}^{n-1} \right) + \frac{37}{24}L\left( \mathbf{Q}^{n-2} \right) - \frac{9}{24}L\left( \mathbf{Q}^{n-3} \right) \right]
\end{eqnarray}
where $L(\mathbf{Q})$ is the residual and $n$ is the current time level. 

The spatial derivatives are discretized using the dispersion-relation-preserving finite difference scheme of Tam and Webb \cite{Tam}. The first derivative at the $l$th node is approximated using $M$ values of $f$ to the right and $N$ values of $f$ to left of the node.
\begin{equation}
\left( \frac{ \partial f }{ \partial x } \right)_l \simeq \frac{1}{\Delta x} \sum_{ j = -N }^{ M } a_j f_{l+j}
\end{equation}
By taking the Fourier transform of the above equation, the coefficients $a_j$ are found by minimizing the integrated error of the difference between the wavenumber of the finite difference scheme and the wavenumber of the Fourier transform of the finite difference scheme. The coefficients $a_j$ are given in Table \ref{stencil_weights}.
\begin{table}
 \begin{center}

  \begin{tabular}{|r|r|r|r|r|r|r|r|} \hline
       Stencil & $a_1=-a_{-1}$ & $a_2=-a_{-2}$ & $a_3=-a_{-3}$ & $a_4=-a_{-4}$   \\\hline
       $DRP$ &  0.77088238 &  -0.16670590 & 0.02084314 & 0 \\ \hline
  \end{tabular}
  \caption{Weights of the centered stencil.}
  \label{stencil_weights}
 \end{center}
\end{table}
To damp out the unwanted high wavenumber waves from the solution, high-order spatial filters, as developed by Kennedy and Carpenter \cite{Kennedy}, are used. 

The Triple Hill's Vortex Synthetic Eddy Method (THV SEM) was applied at the inlet for all cases. No slip boundary condition for velocity and adiabatic condition for temperature are imposed wherever a wall is present and periodic boundary conditions are utilized in homogeneous directions. Standard outflow boundary conditions are applied at the outlet along with a region of artificially increased viscosity immediately upstream of the outlet to damp spurious waves. The dynamic viscosity, $\mu_m$, in the exit region is smoothly increased in the streamwise direction by a weighting function, $w_{\mu}$, an artificial viscosity, $\mu_{BC}$, defined using a Smagorinsky type eddy-viscosity model \cite{Walchshofer}. 
\begin{equation}
\mu\left(x\right) = \mu_m + w_{\mu}\left(x\right)\mu_{BC} = w_{\mu}\left(x\right) \left[\overline{\rho}C_s\Delta^2|\widetilde{S}| \right]
\end{equation}
In practice, $\mu_{BC}$ is equal to the local SGS viscosity calculated by the SGS model in Section \ref{section_SGS}.

\section{Triple Hill's Vortex Synthetic Eddy Method}
Hill's spherical vortex (Hill \cite{Hill}) represents one of the best-known examples of a steady rotational solution to the classical Euler equations, modeling an inviscid incompressible flow. It is characterized by one amplitude, and by one degree of freedom associated with the translation along its axis, which is a fundamental property of the Hill's spherical vortex as a consequence of its definition in relation to a uniform flow. Thus, a synthetic turbulence model that is based on a superposition of eddies representing single Hill's vortices would only have the freedom to reproduce isotropic turbulence. By considering a new vortex structure, one that is composed of three superposed Hill's spherical vortices with the axes perpendicular to each other, two more associated amplitudes are introduced and thus, two more degrees of freedom that can be utilized to match a given Reynolds stress tensor are made available. A synthetic turbulence model consisting of a superposition of such Triple Hill's Vortices (THV) will have enough freedom to reproduce anisotropic turbulence.

\subsection{Hill's Spherical Vortex}
The base component of the proposed synthetic eddy method is the Hill's spherical vortex. It is a steady, axisymmetric solution to the Euler equations for an incompressible flow. Derived from the incompressible Euler equations, the Helmhotz equation for vorticity is
\begin{equation}
\frac{\partial \boldsymbol{\omega}}{\partial t} + (\mathbf{u} \cdot \nabla) \boldsymbol{\omega} = (\boldsymbol{\omega} \cdot \nabla) \mathbf{u}
\end{equation}
where $\boldsymbol{\omega} = (\omega_{r}, \omega_{\theta}, \omega_{z})$. It can be combined with the continuity equation for an incompressible flow,
$
\nabla \cdot \mathbf{u} = 0
$,
and through using the definition of a streamfunction in cylindrical coordinates for an axisymmetric flow,
\begin{equation}\label{stream1}
u_r(r,z) = \frac{1}{r} \frac{\partial \psi}{\partial r}, \hspace{5mm}
u_z(r,z) = -\frac{1}{r} \frac{\partial \psi}{\partial z}
\end{equation}
and the fact that $\omega_{\theta}/r$ is constant along a streamline and only depends on the value of the stream function (i.e. $\partial \left(\omega_{\theta}/r \right)/\partial t = 0$),
\begin{equation}
\frac{\omega_{\theta}}{r} = f(\psi),
\end{equation}
the following equation governing the streamfunction is obtained.
\begin{equation}\label{stream2}
\frac{\partial^2 \psi}{\partial z^2} + \frac{\partial^2 \psi}{\partial r^2} - \frac{1}{r}\frac{\partial \psi}{\partial r} = -r^2 f(\psi)
\end{equation}

Considering the assumption that $\omega_{\theta}/r = f(\psi) = A = const$ inside a sphere of radius $a$, an exact solution to equation (\ref{stream2}) can be found. This solution is known as Hill's spherical vortex \cite{Hill}.
Using the boundary condition at the surface of the sphere, $\psi = 0$, the streamfunction inside the sphere ($r^2 + z^2 < a^2$) is
\begin{equation}
\psi(r,z) = \frac{A r^2}{10} \left(a^2 - z^2 - r^2 \right)
\end{equation}
The streamfunction outside the sphere ($r^2 + z^2 > a^2$),
\begin{equation}
\psi(r,z) = - \frac{u_0 r^2}{2} \left[1 - a^3 \left(r^2 + z^2\right)^{-3/2}\right],
\end{equation}
corresponds to the potential flow around a solid sphere of radius $a$ in a uniform flow of speed $u_0$ in the negative z-axis direction. By matching the two streamfunction solutions at the surface of the sphere, the constant $A$ can be found as
\begin{equation}
A = \frac{15 u_0}{2 a}
\end{equation}
From equation (\ref{stream1}), the velocities in cylindrical coordinates ( $u_r$, $u_z$ ) of the Hill's spherical vortex inside the sphere
\begin{equation}\label{velIN}
u_r(r,z) = \frac{3}{2} u_0 \frac{zr}{a^2}, \hspace{6mm}
u_z(r,z) = \frac{3}{2} u_0 \left( 1 - \frac{z^2 + 2r^2}{a^2} \right)
\end{equation}
and outside the sphere
\begin{equation}\label{velOUT}
u_r(r,z) = \frac{3}{2} u_0 \frac{zr}{a^2} \left( \frac{a^2}{z^2 + r^2} \right)^{5/2}, \hspace{6mm}
u_z(r,z) = u_0 \left[ \left( \frac{a^2}{z^2 + r^2} \right)^{5/2} \frac{2z^2 - r^2}{2a^2} - 1 \right]
\end{equation}
can be found. As a note, the axial velocity outside of the sphere, $u_z$, is presented after the subtraction of the surrounding uniform flow in order to isolate the velocity contribution of the Hill's spherical vortex.

\subsection{Triple Hill's Vortex}
Instead of viewing the fundamental synthetic eddy as a distinct Hill's vortex structure, this method considers a more complex flow structure: the Triple Hill's Vortex (THV). The THV is a superposition of three independent Hill's vortices that all share the same center ($x_0$, $y_0$, $z_0$) and the same outer radius $a$. The rotation axis of each Hill's vortex is orthogonal to the rotation axis of the other two Hill's vortices. In Cartesian coordinates, the rotation axis of one Hill's vortex with amplitude $u_0 = u_0^x$ is oriented in the $x$-direction, and has the associated velocity components given as
\begin{eqnarray}\label{eddyX1}
u^x(x,y,z) &=& u_z(r,z)  \nonumber\\ 
v^x(x,y,z) &=& u_r(r,z) sin(\theta) \\
w^x(x,y,z) &=& u_r(r,z) cos(\theta)\nonumber 
\end{eqnarray}
\begin{equation}
r = \sqrt{y^2 + z^2}, \hspace{6mm} z = x, \hspace{6mm} \theta = arctan \left( \frac{y}{z} \right) \nonumber
\end{equation}
The rotation axis of another Hill's vortex with amplitude $u_0 = u_0^y$ is oriented in the $y$-direction, and has the form
\begin{eqnarray}\label{eddyY1}
u^y(x,y,z) &=& u_r(r,z) cos(\theta) \nonumber\\ 
v^y(x,y,z) &=& u_z(r,z)  \\
w^y(x,y,z) &=& u_r(r,z) sin(\theta)\nonumber 
\end{eqnarray}
\begin{equation}
r = \sqrt{x^2 + z^2}, \hspace{6mm} z = y, \hspace{6mm} \theta = arctan \left( \frac{z}{x} \right) \nonumber
\end{equation}
The rotation axis of the third Hill's vortex with amplitude $u_0 = u_0^z$ is oriented in the $z$-direction, and has the form
\begin{eqnarray}\label{eddyZ1}
u^z(x,y,z) &=& u_r(r,z) sin(\theta) \nonumber\\ 
v^z(x,y,z) &=& u_r(r,z) cos(\theta) \\
w^z(x,y,z) &=& u_z(r,z) \nonumber 
\end{eqnarray}
\begin{equation}
r = \sqrt{x^2 + y^2}, \hspace{6mm} z = z, \hspace{6mm} \theta = arctan \left( \frac{x}{y} \right) \nonumber
\end{equation}
In equations (\ref{eddyX1})-(\ref{eddyZ1}), $r$ and $z$ are the local polar coordinates associated with the vortex (the $z$ coordinate is in the direction of the rotation axis), and $u_r$ and $u_z$ are the velocity components in the local polar coordinate system. The velocity components $u_r$ and $u_z$ of each of the vortices can be found from equations (\ref{velIN}) and {\ref{velOUT}). The superscript signifies the orientation direction of the Hill's vortex rotation axis (for example, $u^y$ is the $x$-component of velocity for the Hill's vortex with the rotation axis oriented in the $y$-direction). Thus, the velocity components for the THV can be obtained by superposition according to 
\begin{eqnarray}\label{velCombined}
\widehat{\widetilde{u}} &=& u^x + u^y + u^z \nonumber \\
\widehat{\widetilde{v}} &=& v^x + v^y + v^z \\
\widehat{\widetilde{w}} &=& w^x + w^y + w^z \nonumber 
\end{eqnarray}
where the tilde ($\widehat{\widetilde{\quad}}$) signifies the entire THV. 

\subsection{Divergence of a Triple Hill's Vortex}
Consider the divergence of the velocity field for a single THV, $\mathbf{\widehat{\widetilde{u}}}$.
\begin{eqnarray}
\nabla \cdot \mathbf{\widehat{\widetilde{u}}} &=& \frac{\partial \widehat{\widetilde{u}}}{\partial x} + \frac{\partial \widehat{\widetilde{v}}}{\partial y} + \frac{\partial \widehat{\widetilde{w}}}{\partial z} \nonumber \\
&=& \frac{\partial \left(u^x + u^y + u^z\right)}{\partial x} + \frac{\partial \left(v^x + v^y + v^z\right)}{\partial y} + \frac{\partial \left(w^x + w^y + w^z\right)}{\partial z}  \\
&=& \left(\frac{\partial u^x}{\partial x} + \frac{\partial v^x}{\partial y} + \frac{\partial w^x}{\partial z} \right)
  + \left(\frac{\partial u^y}{\partial x} + \frac{\partial v^y}{\partial y} + \frac{\partial w^y}{\partial z} \right)
  + \left(\frac{\partial u^z}{\partial x} + \frac{\partial v^z}{\partial y} + \frac{\partial w^z}{\partial z} \right) \nonumber \\
&=& \left( \nabla \cdot \mathbf{u}^x \right) + \left( \nabla \cdot \mathbf{u}^y \right) + \left( \nabla \cdot \mathbf{u}^z \right) \nonumber 
\end{eqnarray} 
where, for example, $u^x$, $v^x$ and $w^x$ are the velocity components in the local coordinate system that is associated with the vortex with its axis aligned with the $x$ axis (of the global coordinate system). Therefore, since each Hill's vortex satisfies the divergence-free condition in its local coordinate system, 
\begin{equation}
\nabla \cdot \mathbf{\widehat{\widetilde{u}}} = 0 
\end{equation} 
the THV is divergence-free.

\subsection{Convection of a Triple Hill's Vortex}
The THV's are convected through the inlet using Taylor's frozen turbulence hypothesis 
\begin{eqnarray}\label{convection1}
x' = (x - x_0) - U( t - t0 ) \nonumber \\
y' = (y - y_0) - V( t - t0 ) \\
z' = (z - z_0) - W( t - t0 ) \nonumber
\end{eqnarray}
 where ($U$,$V$,$W$) are the mean velocity components and ($x_0$,$y_0$,$z_0$,$t_0$) are the space and time coordinates of the center of the THV. The divergence of velocity for a single THV is only identically zero when ($U$,$V$,$W$) are constant over the entire THV. For spatially varying inflow velocity profiles, such as in channel flow, the mean inflow velocity vary continuously across the THV. So, the following assumption was made: the ($U$,$V$,$W$) for a single THV are calculated at the center of the THV and applied over the entire THV.
\begin{eqnarray}\label{convection2}
x' = (x - x_0) - U(x_0,y_0,z_0) ( t - t0 ) \nonumber \\
y' = (y - y_0) - V(x_0,y_0,z_0) ( t - t0 ) \\
z' = (z - z_0) - W(x_0,y_0,z_0) ( t - t0 ) \nonumber
\end{eqnarray}

\subsection{Inflow Velocity}
The calculation of the THV Synthetic Eddy Method is similar to the method proposed by Jarrin et al. \cite{Jarrin_1}, except the matching between the imposed and given Reynolds stress tensors is performed differently, in order to maintain the divergence-free condition. Also, the previous methods used Gaussian and tent functions to define the shape of these eddies, which may not reproduce realistic flow fields. Since the synthetic eddies (THV) proposed in this model are composed of Hill's vortices, which are actual solutions to the Euler equation, it is expected that they will resemble more realistic turbulent flow conditions. The cases presented in Section \ref{results} will investigate whether this expectation is met.  The turbulent inflow is viewed as a collection of eddies added to a mean flow. In the following, the velocities in equations (\ref{eddyX1}), (\ref{eddyY1}), and (\ref{eddyZ1}) are assumed to be products of a constant amplitude $\left( u_0^x, u_0^y, u_0^z \right)$ multiplied by a shape function $\left( f_i^x, f_i^y, f_i^z \right)$, 

\begin{equation}\label{shape}
\left[
\begin{array}{c}
u^x   \\
v^x   \\
w^x   
\end{array}
\right] = \left[
\begin{array}{c}
u_0^x f_u^x \\
u_0^x f_v^x \\
u_0^x f_w^x 
\end{array}
\right], \hspace{6mm}
\left[
\begin{array}{c}
u^y   \\
v^y   \\
w^y   
\end{array}
\right] = \left[
\begin{array}{c}
u_0^y f_u^y \\
u_0^y f_v^y \\
u_0^y f_w^y 
\end{array}
\right], \hspace{6mm}
\left[
\begin{array}{c}
u^z   \\
v^z   \\
w^z   
\end{array}
\right] = \left[
\begin{array}{c}
u_0^z f_u^z \\
u_0^z f_v^z \\
u_0^z f_w^z 
\end{array}
\right]
\end{equation}
where, instead of using Gaussian shape functions or other simple functions (as in Jarrin et al. \cite{Jarrin_1} and Poletto \cite{Poletto_1}), THV shape functions are utilized (extracted from equations (\ref{velIN}) and (\ref{velOUT})). The three amplitudes associated with the THV are the means by which the Reynolds stress tensor components of a prescribed inflow velocity are matched. 
The inflow velocity components, ($u_{in}$, $v_{in}$, $w_{in}$), are composed of the mean base flow and the fluctuating components 
\begin{eqnarray}\label{inflow}
u_{in} &=& U + \sum_{j=1}^{N} \widehat{\widetilde{u}}_j \nonumber \\
v_{in} &=& V + \sum_{j=1}^{N} \widehat{\widetilde{v}}_j \\
w_{in} &=& W + \sum_{j=1}^{N} \widehat{\widetilde{w}}_j \nonumber
\end{eqnarray} 
where ($U$,$V$,$W$) is the mean velocity vector, and ($\widehat{\widetilde{u}}_j$,$\widehat{\widetilde{v}}_j$,$\widehat{\widetilde{w}}_j$) is the velocity vector of the $j$th THV. Each THV has an independent random center, radius, and amplitude. 

\subsection{Determination of the Amplitudes} \label{section_amplitude}
In the original SEM of Jarrin et al. \cite{Jarrin_1}, the fluctuating velocity field is rescaled using the Cholesky decomposition of the Reynolds stress tensor in order to ensure reproduction of the desired Reynolds stresses. Since those synthetic eddies already violate the divergence-free condition, the fact that rescaling using the Cholesky decomposition generally violates the divergence-free condition is of little consequence. Because the THV is divergence-free, the amplitudes of the THV's need to be calculated using a different method in order to match the desired Reynolds stresses and also preserve the divergence-free nature of the synthetic field.

Smirnov et al. \cite{Smirnov} proposed a method, which was later expanded upon by Davidson and Bilson \cite{Davidson}, to match the Reynolds stresses for anisotropic turbulence. The idea behind the method is to find a local reference system where the normal Reynolds stress terms are non-zero, while the off-diagonal Reynolds stress terms are zero, calculate the associated amplitudes, and then transform the signal back to the global reference system. To be able to apply this method in the current THV framework, the creation of the THV's has to take place in the local principal-axis reference system. This begins by calculating the principal-axis Reynolds stresses and eigenvectors of the local Reynolds stress tensor. Examining the principal-axis Reynolds stresses for a single stream of THV's can give insight into how to set each of the principal-axis amplitudes, ($u_0^{x,p}$, $u_0^{y,p}$, $u_0^{z,p}$), where superscript $^p$ denotes variables in the principal-axis coordinate system. Since the velocity components are at a maximum, consider the principal-axis Reynolds stresses at the center of a single stream of THV's. A single stream of THV's is defined as an infinite number of identical THV's whose centers pass through the same point in space on the inflow  plane. The frequency with which the THV's pass through the inflow plane is such that the preceeding and subsequent THV's do not interact with the THV currently at the inlet.
\begin{eqnarray}\label{ReStress1}
\left<\widehat{\widetilde{u}}\widehat{\widetilde{u}}\right>^p &=& \left<\left(u_0^{x,p}\right)^2 \left(f_u^{x,p}\right)^2 \right> + \left<\left(u_0^{y,p}\right)^2 \left(f_u^{y,p}\right)^2 \right> + \left<\left(u_0^{z,p}\right)^2 \left(f_u^{z,p}\right)^2 \right> \nonumber \\
\left<\widehat{\widetilde{v}}\widehat{\widetilde{v}}\right>^p &=& \left<\left(u_0^{x,p}\right)^2 \left(f_v^{x,p}\right)^2 \right> + \left<\left(u_0^{y,p}\right)^2 \left(f_v^{y,p}\right)^2 \right> + \left<\left(u_0^{z,p}\right)^2 \left(f_v^{z,p}\right)^2 \right> \\
\left<\widehat{\widetilde{w}}\widehat{\widetilde{w}}\right>^p &=& \left<\left(u_0^{x,p}\right)^2 \left(f_w^{x,p}\right)^2 \right> + \left<\left(u_0^{y,p}\right)^2 \left(f_w^{y,p}\right)^2 \right> + \left<\left(u_0^{z,p}\right)^2 \left(f_w^{z,p}\right)^2 \right> \nonumber
\end{eqnarray}
\begin{equation}
\left<\widehat{\widetilde{u}}\widehat{\widetilde{v}}\right>^p = \left<\widehat{\widetilde{u}}\widehat{\widetilde{w}}\right>^p = \left<\widehat{\widetilde{v}}\widehat{\widetilde{w}}\right>^p = 0 \nonumber
\end{equation} 
Let the amplitudes in the principal-axis coordinate system associated with a THV be defined as a product of a constant amplitude and a random number,
\begin{eqnarray}\label{epsilon1}
\left[
\begin{array}{c}
u_0^{x,p} \\
u_0^{y,p} \\
u_0^{z,p} 
\end{array}
\right] = \left[
\begin{array}{c}
\epsilon^x\ \hat{u}_0^{x,p} \\
\epsilon^y\ \hat{u}_0^{y,p} \\
\epsilon^z\ \hat{u}_0^{z,p} 
\end{array}
\right]
\end{eqnarray}
where $\epsilon^x$, $\epsilon^y$, and $\epsilon^z$ are independent random numbers such that $\epsilon = \pm 1$, $\left< \epsilon \right> = 0$, $\left< \epsilon^2 \right> = 1$, and $\left< \epsilon^x \epsilon^y \right> = \left< \epsilon^x \epsilon^z \right> = \left< \epsilon^y \epsilon^z \right> = 0$ .
Since the amplitudes and the constant shape functions are independent, after inserting equation (\ref{epsilon1}) and using the properties of $\epsilon$, equation (\ref{ReStress1}) becomes
\begin{eqnarray}\label{ReStress2}
\left[
\begin{array}{c}
\left<\widehat{\widetilde{u}}\widehat{\widetilde{u}}\right>^p \\
\left<\widehat{\widetilde{v}}\widehat{\widetilde{v}}\right>^p \\
\left<\widehat{\widetilde{w}}\widehat{\widetilde{w}}\right>^p
\end{array}
\right] = \left[
\begin{array}{ccc}
\left(f_u^{x,p}\right)^2  \left(f_u^{y,p}\right)^2  \left(f_u^{z,p}\right)^2  \\
\left(f_v^{x,p}\right)^2  \left(f_v^{y,p}\right)^2  \left(f_v^{z,p}\right)^2  \\
\left(f_w^{x,p}\right)^2  \left(f_w^{y,p}\right)^2  \left(f_w^{z,p}\right)^2 
\end{array}
\right] \left[
\begin{array}{c}
\left(\hat{u}_0^{x,p}\right)^2 \\
\left(\hat{u}_0^{y,p}\right)^2 \\
\left(\hat{u}_0^{z,p}\right)^2 
\end{array}
\right] = F_{uvw}^{xyz} \left[
\begin{array}{c}
\left(\hat{u}_0^{x,p}\right)^2 \\
\left(\hat{u}_0^{y,p}\right)^2 \\
\left(\hat{u}_0^{z,p}\right)^2 
\end{array}
\right]
\end{eqnarray}
$F_{uvw}^{xyz}$ is constant and invertible. Thus, the amplitudes of a THV can be calculated from the given principal-axis Reynolds stresses.
By combining the amplitudes calculated in equation (\ref{ReStress2}) with equations (\ref{epsilon1}) and (\ref{shape}),
\begin{equation}\label{shape_principal}
\left[
\begin{array}{c}
u^{x,p}   \\
v^{x,p}   \\
w^{x,p}   
\end{array}
\right] = \left[
\begin{array}{c}
\epsilon^x\ \hat{u}_0^{x,p} f_u^{x,p} \\
\epsilon^x\ \hat{u}_0^{x,p} f_v^{x,p} \\
\epsilon^x\ \hat{u}_0^{x,p} f_w^{x,p} 
\end{array}
\right], 
\left[
\begin{array}{c}
u^{y,p}   \\
v^{y,p}   \\
w^{y,p}   
\end{array}
\right] = \left[
\begin{array}{c}
\epsilon^y\ \hat{u}_0^{y,p} f_u^{y,p} \\
\epsilon^y\ \hat{u}_0^{y,p} f_v^{y,p} \\
\epsilon^y\ \hat{u}_0^{y,p} f_w^{y,p} 
\end{array}
\right], 
\left[
\begin{array}{c}
u^{z,p}   \\
v^{z,p}   \\
w^{z,p}   
\end{array}
\right] = \left[
\begin{array}{c}
\epsilon^z\ \hat{u}_0^{z,p} f_u^{z,p} \\
\epsilon^z\ \hat{u}_0^{z,p} f_v^{z,p} \\
\epsilon^z\ \hat{u}_0^{z,p} f_w^{z,p} 
\end{array}
\right]
\end{equation}
and then inserting the result into equation (\ref{velCombined}), the velocity contribution of a single THV in the local principal-axis reference system can be found as 
\begin{eqnarray}\label{velCombined_principal}
\widehat{\widetilde{u}}^p &=& u^{x,p} + u^{y,p} + u^{z,p} \nonumber \\
\widehat{\widetilde{v}}^p &=& v^{x,p} + v^{y,p} + v^{z,p} \\
\widehat{\widetilde{w}}^p &=& w^{x,p} + w^{y,p} + w^{z,p} \nonumber 
\end{eqnarray}
The last step is to transform the velocities from the local principal-axis reference system to the global reference system using the transformation matrix created from the eigenvectors of the local Reynolds stress tensor, $T_{p}^{G}$. 
\begin{eqnarray}\label{transform_p_G}
\left[
\begin{array}{c}
\widehat{\widetilde{u}} \\
\widehat{\widetilde{v}} \\
\widehat{\widetilde{w}} 
\end{array}
\right] &=& T_{p}^{G} \left[
\begin{array}{c}
\widehat{\widetilde{u}}^p \\
\widehat{\widetilde{v}}^p \\
\widehat{\widetilde{w}}^p 
\end{array}
\right]
\end{eqnarray}
This is the velocity contribution of a single THV in the global reference system that is combined with all the other THV's in equation (\ref{inflow}) to create the imposed inflow.

\subsection{Modification of the Target Reynolds Stresses} \label{target_reynolds_stress_modification}
As is defined in Section \ref{section_amplitude}, the imposed Reynolds stress tensor at any point in space is reproduced by a single stream of THV's moving through that point in time. To account for that fact that the imposed turbulent inflow is a superposition of many THV's over a range of different sizes, the method for determining the amplitudes of each THV needs to include the influence of the THV's currently at the inlet when a new THV is created in order to ensure that Reynolds stress matching is recovered. The target Reynolds stress modification procedure is proposed to address these issue. 

Consider the inlet plane shown in Figure \ref{scaling_inlet}. The black circles represent THV's that are currently passing through the inlet and the red circle represents a new THV that is about to be created. If the surrounding THV's did not exist, the calculated amplitudes of the new THV ensure Reynolds stress matching at the center of the THV. Since the surrounding THV's already exist and are influencing the flow field at the center of the new THV, the amplitudes of the new THV need to be modified to account for the contributions of its neighbors. 

\begin{figure}
 \begin{center}
  \begin{tikzpicture}

     \draw[thick] (0,0) rectangle (8,8);
     \draw[red] (4,4) circle (30pt);
     \draw[fill=black] (4,4) circle (0.5pt)node[anchor=east] {new};
     
     \draw (4,4.75) circle (90pt);
     \draw[fill=black] (4,4.75) circle (0.5pt) node[anchor=west] {$1$};
     
     \draw (4.75,3) circle (80pt);
     \draw[fill=black] (4.75,3) circle (0.5pt) node[anchor=north] {$2$};
     
     \draw (3,2.5) circle (60pt);
     \draw[fill=black] (3,2.5) circle (0.5pt) node[anchor=north] {$3$};
     
     \draw (1.5,6) circle (35pt);
     \draw[fill=black] (1.5,6) circle (0.5pt) node[anchor=north] {$4$};
     
     \draw (7,5.5) circle (25pt);
     \draw[fill=black] (7,5.5) circle (0.5pt) node[anchor=north] {$5$};
     
     \draw (7,1) circle (20pt);
     \draw[fill=black] (7,1) circle (0.5pt) node[anchor=north] {$6$};
     
     \draw (1.5,1.5) circle (40pt);
     \draw[fill=black] (1.5,1.5) circle (0.5pt) node[anchor=north] {$8$};
       
     \draw[<->] (4,4) -- (4,4.75);
     \draw[<->] (4,4) -- (4.75,3);
     \draw[<->] (4,4) -- (3,2.5);
     
  \end{tikzpicture}
 \end{center}
  \caption{ An example of inlet plane when a new THV is being created: black) The current THV's being convected through the inflow plane; red) The random location on the inflow plane where a new THV is being created. }
 \label{scaling_inlet}
\end{figure}
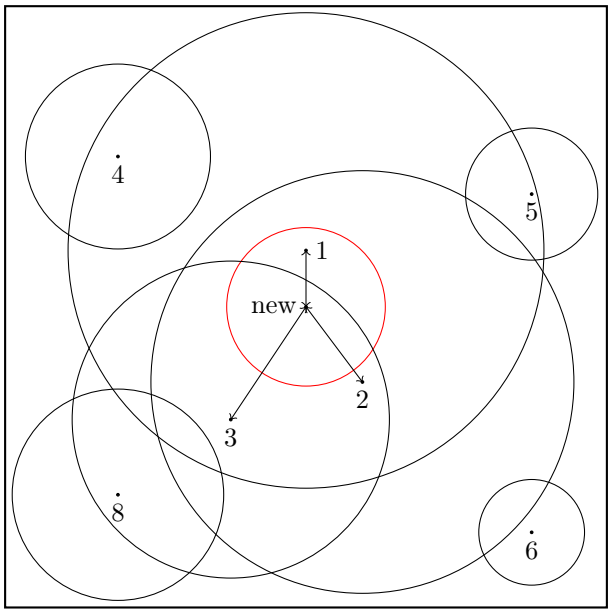

Looking at equation (\ref{velOUT}), the velocity components for a Hill's vortex approach zero quickly when moving away from the surface of the vortex. Thus, only the contribution of THV's that encircle the center of the new THV will be considered (the center of THV new falls within the boundaries of THV 1,2, and 3). In the global coordinate system, the Reynolds stresses at the point on the inlet where the center of the new THV will pass through are defined as 
\begin{eqnarray}
\langle u'_i u'_j \rangle_{existing} &=& \left\langle \left( \sum_{m=1}^{M_s} \left( \widetilde{u}_i \right)_m \right) \left( \sum_{n=1}^{N_s} \left( \widehat{\widetilde{u}}_j \right)_n \right) \right\rangle \\
&=& \sum_{m=1}^{M_s} \sum_{n=1}^{N_s} \langle \left( \widehat{\widetilde{u}}_i \right)_m \left( \widehat{\widetilde{u}}_j \right)_n \rangle
\end{eqnarray}
where $M_s$ and $N_s$ are both equal to  the number of encircling THV's. Since the random numbers associated with each THV are independent,
\begin{eqnarray}
\langle \left( \widehat{\widetilde{u}}_i \right)_m \left( \widehat{\widetilde{u}}_j \right)_n \rangle = 0 \quad \text{for} \quad m \neq n
\end{eqnarray}
Thus, the Reynolds stress contribution from the surrounding THV's is
\begin{eqnarray} \label{THV_contribution}
\langle u'_i u'_j \rangle_{existing} = \sum_{n=1}^{N_s} \langle \widehat{\widetilde{u}}_i \widehat{\widetilde{u}}_j \rangle_n
\end{eqnarray}
where $N_s$ is again the number of encircling THV's. This contribution Reynolds stress tensor is then used to modify the given Reynolds stress tensor that is being matched.
\begin{eqnarray} \label{THV_target}
\langle u'_i u'_j \rangle_{target} = \langle u'_i u'_j \rangle_{given} - \langle u'_i u'_j \rangle_{existing}
\end{eqnarray}
The target Reynolds stress tensor can then be used in the THV creation framework to determine the amplitudes of the new THV. As a note, if any of the eigenvalues of the target Reynolds stress tensor are negative, that means that the given Reynolds stresses are already reproduced by the present THV's and the new THV does not need to be created at that point on the inlet.

\subsubsection{Reynolds Stresses for a Single Stream of THV} \label{single_THV_Re_stress}
The Reynolds stress tensors of the individual THV's needed by equation (\ref{THV_contribution}) can be found by first considering the velocity components of a single THV in the global coordinate system. 
\begin{eqnarray}\label{THV_velocity_global}
\widehat{\widetilde{u}}_i = u_0^x f_i^x + u_0^2 f_i^y + u_0^3 f_i^z
\end{eqnarray} 
$\left( f_i^x, f_i^y, f_i^z \right)$ are the shape functions for the three Hill's vortices associated with the $\widehat{\widetilde{u}}_i$ velocity component. $\left( u_0^x, u_0^y, u_0^z \right)$ are the randomized amplitudes in the global coordinate system and are defined as
\begin{eqnarray}\label{global_amplitudes}
\left[
\begin{array}{c}
u_0^x \\
u_0^y \\
u_0^z
\end{array}
\right] &=& T_p^G \left[
\begin{array}{c}
\epsilon^x\thinspace \hat{u}_0^{x,p} \\ 
\epsilon^y\thinspace \hat{u}_0^{y,p} \\
\epsilon^z\thinspace \hat{u}_0^{z,p}
\end{array}
\right]
\end{eqnarray}
where $T_p^G$ is the eigenvector transformation matrix, $\left( \hat{u}_0^{x,p},\hat{u}_0^{y,p},\hat{u}_0^{z,p} \right)$ are the constant amplitudes in the principal coordinate system, and $\left( \epsilon^x, \epsilon^y, \epsilon^z \right)$ are independent random numbers that have the following properties
\begin{eqnarray}\label{epsilon_properties}
&\epsilon^x = \pm 1 \quad;\quad \epsilon^y = \pm 1 \quad;\quad \epsilon^z = \pm 1 \\ \nonumber
&\langle \epsilon^x \rangle = \langle \epsilon^y \rangle = \langle \epsilon^z \rangle = 0 \\ \nonumber
&\langle \left(\epsilon^x\right)^2 \rangle = \langle \left(\epsilon^y\right)^2 \rangle = \langle \left(\epsilon^z\right)^2 \rangle = 1 \\ \nonumber
&\langle \epsilon^x\epsilon^y \rangle = \langle \epsilon^x\epsilon^z \rangle = \langle \epsilon^y\epsilon^z \rangle = 0 \nonumber
\end{eqnarray}
By taking the time-average of the products of the velocity components in equation (\ref{THV_velocity_global}) and recognizing that the amplitudes and shape functions are independent from each other, expressions can be found for the Reynolds stresses of an infinite stream of THV's being convected through the same location at the inlet. 
\begin{eqnarray}\label{THV_Re_stress}
\langle \widehat{\widetilde{u}}_i \widehat{\widetilde{u}}_j \rangle = 
\langle \left( u_0^x \right)^2 \rangle \langle f_i^x f_j^x \rangle &+&
\langle \left( u_0^y \right)^2 \rangle \langle f_i^y f_j^y \rangle 
+ \langle \left( u_0^z \right)^2 \rangle \langle f_i^z f_j^z \rangle \\ \nonumber
&+& \left( \langle f_i^x f_j^y \rangle + \langle f_j^x f_i^y \rangle \right) \langle u_0^x u_0^y \rangle  
+ \left( \langle f_i^x f_j^z \rangle + \langle f_j^x f_i^z \rangle \right) \langle u_0^x u_0^z \rangle \\
&+& \left( \langle f_i^y f_j^z \rangle + \langle f_j^y f_i^z \rangle \right) \langle u_0^y u_0^z \rangle \nonumber
\end{eqnarray}

The time-average of the products of the shape functions can be solved for analytically by assuming the time period for a single THV to flow through the inlet is 
\begin{eqnarray}
T_{in} = \frac{4\thinspace a}{U}
\end{eqnarray}
where $a$ is the radius of the THV and $U$ is the mean velocity at the center of the THV.

Through the multiplying of the global amplitudes in equation (\ref{global_amplitudes}), taking the time-average of the resulting products, recognizing that the eigenvector transformation matrix and the principal coordinate system amplitudes are all constant, and finally inserting the properties of the random numbers seen in equation (\ref{epsilon_properties}); the time-averaged global coordinate system amplitude correlations can be determined.
\begin{eqnarray}\label{amplitude_average}
\langle u_0^i u_0^j \rangle = \left(t_p^G\right)_{i1} \left(t_p^G\right)_{j1} \left( \hat{u}_0^{x,p} \right)^2 
+ \left(t_p^G\right)_{i2} \left(t_p^G\right)_{j2} \left( \hat{u}_0^{y,p} \right)^2 
+ \left(t_p^G\right)_{i3} \left(t_p^G\right)_{j3} \left( \hat{u}_0^{z,p} \right)^2
\end{eqnarray}
$t_p^G$ are the individual elements of the eigenvector transformation matrix, $T_p^G$.

\subsection{Generations of Triple Hill's Vortices}
To be able to recreate turbulent flow, the Reynolds stress tensor is not the only quantity that needs to be matched. Turbulent structures occur over a wide range of length scales, and the population varies over a  length scale distribution which can be taken from the turbulent kinetic energy spectrum. Thus, the number of larger structures is less than the number of medium sized structures, which is less than the number of smaller structures. This type of distribution also needs to be replicated to make sure the space is fully covered by structures in the whole range of wavenumbers. The minimum, $a_{min}$, and maximum, $a_{max}$, radii of the THV allowed are dictated by the grid resolution and the integral length scale or the domain size, respectively. As shown in Jarrin et al. \cite{Jarrin_2}, the overall number of imposed THV's are calculated as follows,
\begin{eqnarray}\label{eddy_number}
N &=& C\frac{A_{in}}{A_{THV}}
\end{eqnarray}
where $N$ is the total number of THV's, $A_{in}$ is the area of the inlet plane, $A_{THV}$ is the projected area of a THV onto the inlet plane, and $C$ is a proportionality constant. This number of THV's will ensure that the inflow plane is fully populated with THV's at any moment in time. Jarrin et al. \cite{Jarrin_2} recommends that $C = 1$ to minimize the computational overhead required with any unnecessary THV's.
Given the limits for the smallest and the largest eddy, the idea of defining multiple generations of THV's is introduced here. The range of radii is divided discretely into several subranges, or generations. Based on a given distribution, the number of THV's in each generation is set. Table \ref{THV_gen} shows an example of five generations of THV's, where $a_{min} < a_4 < a_3 < a_2 < a_1 < a_{max}$ and $N = N_1 + N_2 + N_3 + N_4 + N_5$

\begin{table}[htpb]
 \begin{center}
  \begin{tabular}{|c|c|c|c|c|c|} \hline
       Generation  & Range of Radii & Number of THV's\\\hline
       $1$ &  $a_1 < a < a_{max}$  &  $N_1$  \\    \hline
       $2$ &  $a_2 < a < a_1$      &  $N_2$  \\    \hline
       $3$ &  $a_3 < a < a_2$      &  $N_3$  \\    \hline
       $4$ &  $a_4 < a < a_3$      &  $N_4$  \\    \hline
       $5$ &  $a_{min} < a < a_4$  &  $N_5$  \\    
       \hline
  \end{tabular}
  \caption{Example of five generations of THV's}
  \label{THV_gen}
 \end{center}
\end{table}

The multiple generations of THV's ensure that the inlet is covered with enough multiscale structures. The range of locations for these generations can also be restricted, for example, to cluster smaller THV's near a wall for wall-bounded flows.

Each generation of THV can either match all or only a portion of the overall Reynolds stress tensor. The given Reynolds stress tensor in equation (\ref{THV_target}) is multiplied by a factor, $b_m$, based on the ratio of turbulent kinetic energy (TKE) contained in that specific generation,$TKE_m$, to the total turbulent kinetic energy, $TKE$.
\begin{eqnarray}
b_m = \frac{TKE_m}{TKE} \quad\quad ; \quad\quad \sum_{m=1}^{M} b_m = 1
\end{eqnarray}
$M$ is the total number of THV generations and $m$ represents the current generation. Thus, the equation for the target Reynolds stresses for a particular THV, equation (\ref{THV_target}), is modified as follows.
\begin{eqnarray}
\langle u'_i u'_j \rangle_{target} = b_m\langle u'_i u'_j \rangle_{given} - \langle u'_i u'_j \rangle_{existing}
\end{eqnarray}
The Reynolds stress contribution from the surrounding THV's, $\langle u'_i u'_j \rangle_{existing}$, is only calculated over the THV's in the same generation as the THV being created.
\begin{eqnarray}
\langle u'_i u'_j \rangle_{existing} = \sum_{n=1}^{N_M} \langle \widehat{\widetilde{u}}_i \widehat{\widetilde{u}}_j \rangle_n
\end{eqnarray}
$N_m$ is the number of encircling THV's in the same generation. By only matching a portion of the TKE, THV's from multiple generations can exist on top of each other. This enables energy to be contributed over a wider range of scales. The number of generations, the radii bounds of the generations, and the factor, $b_m$ all provide an element of control over how the TKE is distributed over the resolvable lengthscales. More generations and tighter radii bounds for the generations allow for a finer tuning of the spectra.

\subsection{Near-wall THV Stretching} \label{section_stretch}
A spherical Triple Hill's Vortex created near a wall does not agree with the observed elongated flow structures observed in the near-wall region. A stretching operation is introduced to produce angled, elongated THV's in the near-wall region consistent with the observations of Jeong et al. \cite{Jeong} and Sibilla and Beretta \cite{Sibilla}. The physical coordinates system associated with the center of a near-wall THV is stretched according to a streamwise length scale, rotated about the spanwise coordinate axis according to a inclination angle, and rotated about the vertical coordinate axis according to a tilting angle. Figure \ref{THV_stretch} shows a diagram illustrating the stretched and rotated THV.
\begin{equation} \label{stretching}
\left[
	\begin{array}{c}
		x'_s \\
		y'_s \\
		z'_s 
	\end{array}
\right] = \left[
	\begin{array}{ccc}
		\frac{1}{2}\frac{l_x}{2a}	& 0	& 0 \\
		0							& 1	& 0	\\
		0							& 0	& 1	
	\end{array}
\right] \left[
	\begin{array}{ccc}
		cos(\alpha)	& -sin(\alpha)	& 0	\\
		sin(\alpha)	& cos(\alpha)	& 0	\\
		0			& 0				& 1	
	\end{array}
\right] \left[
	\begin{array}{ccc}
		cos(\beta)	& 0	& -sin(\beta)	\\
		0			& 1	& 0				\\
		sin(\beta)	& 0	& cos(\beta)
	\end{array}
\right] \left[
	\begin{array}{c}
		x' \\
		y' \\
		z'
	\end{array}
\right]
\end{equation}
$(x',y',z')$ are the physical coordinates associated with the center of a THV and $(x_s',y_s',z_s')$ are the stretched and rotated coordinates that are then used to calculate the velocity components of the THV. $l_x/2a$ is the length to diameter ratio and it is multiplied by $1/2$ in equation (\ref{stretching}) because the spherical THV is stretched equally in the positive and negative streamwise direction. The inclination angle is $\alpha$ and $\beta$ is the tilting angle. Jeong et al. \cite{Jeong} found that the inclination angle was constant and that the tilting angle varied between $-\beta_{max}$ and $\beta_{max}$ depending on the local vertical component of vorticity. Here, 
\begin{equation}
\beta = \epsilon^\beta \beta_{max}
\end{equation}
where $\epsilon^\beta$ is a random number defined such that $\langle \epsilon^\beta \rangle = 0$ and $\langle \epsilon^\beta \thinspace^2 \rangle = 1$. The constant stretching and rotation procedure allows for the continued satisfaction of the divergence-free condition for a THV. 

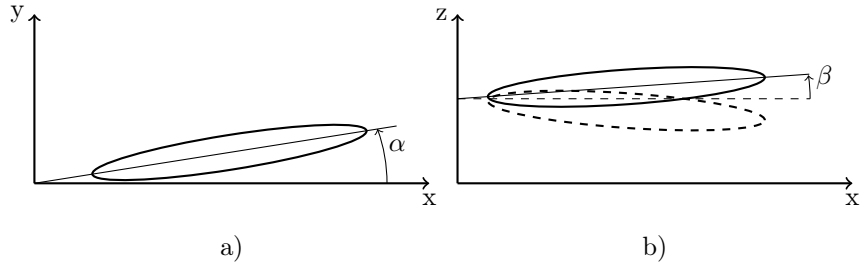
\begin{figure}
 \begin{center}
  \begin{tikzpicture}

     \draw (2.625,-0.5625) node[anchor=north] {a)};
     \draw[->,thick] (0,0) -- (5.25,0) node[anchor=north] {x};
     \draw[->,thick] (0,0) -- (0,2.25) node[anchor=east] {y};
     \draw[rotate around={9:(0,0)},thick] (2.625,0) ellipse (52.5pt and 6.5625pt);
     \draw[rotate around={9:(0,0)}] (0,0) -- (4.875,0);
     \draw[->] (4.6875,0) arc (0:20:60pt);
     \draw (5.0625,0.5) node[anchor=east] {$\alpha$};
     
     \draw (8.25,-0.5625) node[anchor=north] {b)};
     \draw[->,thick] (5.625,0) -- (10.875,0) node[anchor=north] {x};
     \draw[->,thick] (5.625,0) -- (5.625,2.25) node[anchor=east] {z};
     \draw[rotate around={4:(5.625,1.125)},thick] (7.875,1.125) ellipse (52.5pt and 6.5625pt);
     \draw[rotate around={-4:(5.625,1.125)},thick,dashed] (7.875,1.125) ellipse (52.5pt and 6.5625pt);
     \draw[rotate around={4:(5.625,1.125)}] (5.625,1.125) -- (10.3125,1.125);
     \draw[dashed] (5.625,1.125) -- (10.3125,1.125);
     \draw[->] (10.3125,1.125) arc (0:8.5:60pt);
     \draw (10.5,1.125) node[anchor=south] {$\beta$};

  \end{tikzpicture}
 \end{center}
  \caption{A diagram of the stretched THV imposed in the near-wall region: a) inclination angle; b) tilting angle.}
 \label{THV_stretch}
\end{figure}

\section{Results} \label{results}

Three test cases were considered: convection of an isolated Triple Hill's Vortex, homogeneous isotropic turbulence, and turbulent channel flow.

\subsection{Single Triple Hill's Vortex}
The convection of a single Triple Hill's Vortex was investigated. The THV was generated at the inlet plane and convected downstream by a uniform mean flow. The dimensions of the domain were $12a \times 6a \times 6a$ in the streamwise, vertical, and spanwise directions and it was discretized using a uniform Cartesian grid with $120 \times 60 \times 60$ grid points. Far field boundary conditions were used on the vertical and spanwise boundaries.

Contours of the velocity magnitude on planes through the center of a single convected THV are shown in Figure \ref{THV_vel_mag}. The contours clearly shows the spherical, and thus symmetric, nature of the THV.  The xy- and xz-planes also show the stretching of the THV by the convecting flow. 

\begin{figure}
 \begin{center}
  \mbox{
      \includegraphics[width=9.cm]{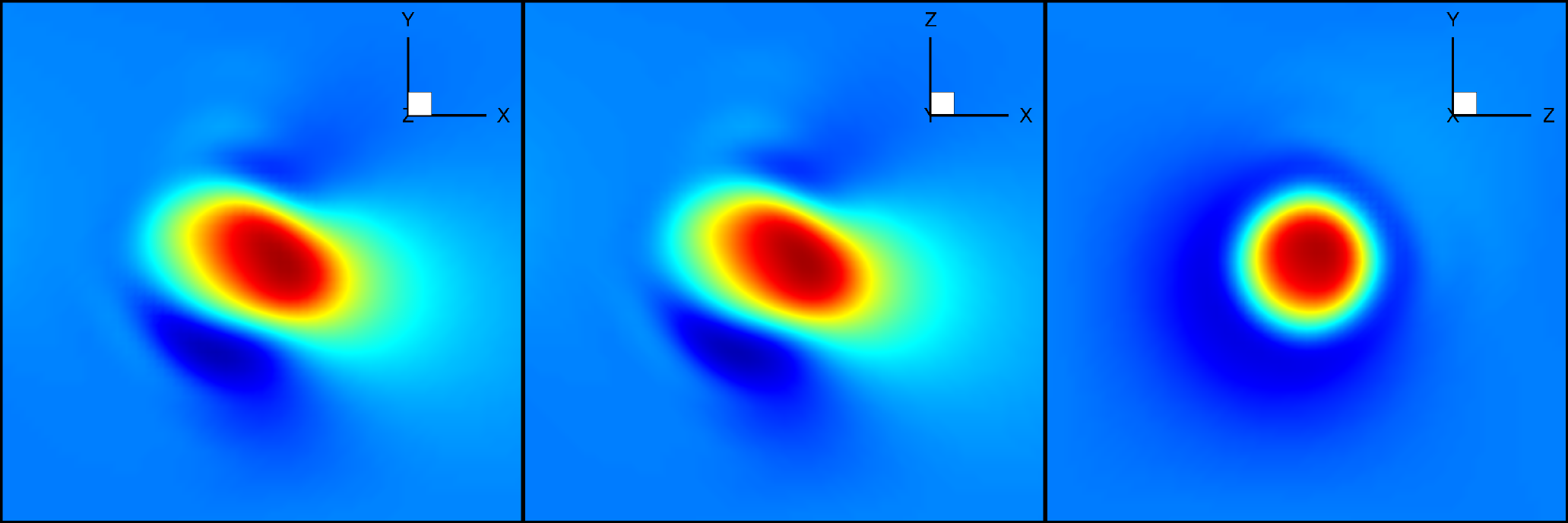}
         }
      \\ xy-plane) \hspace{15mm} xz-plane)  \hspace{15mm} yz-plane) 
 \end{center}
  \caption{{Contours of the velocity magnitude on xy\thinspace(left), xz\thinspace(middle), and yz\thinspace(right) planes through the center of a THV.}}
  \label{THV_vel_mag}
\end{figure}

Figure \ref{THV_time1} depicts the generation in time of a single THV from the inlet. Each of the frames is a constant time step apart. In the first frame, the THV can be seen just beginning to emerge from the inlet. Notice that there are no spurious waves originating from the front of the THV. Then, moving forward in time, the next two frames show the THV as half of it is generated and then as it just leaves the inlet. In the last frame, the THV has fully released from the inlet. Notice that the THV passed through the inflow boundary cleanly, with no spurious waves originating from the back of the THV, which is an indication that the divergence-free condition is satisfied (otherwise, spurious waves may be generated in the downstream of the eddy). In the synthetic turbulence model, analyzed next, when a THV is released from the inlet, it is no longer acted upon by the THV SEM and the generation process begins again with a new THV at a new random location.

\begin{figure}
 \begin{center}
  \mbox{
      \includegraphics[width=12.cm]{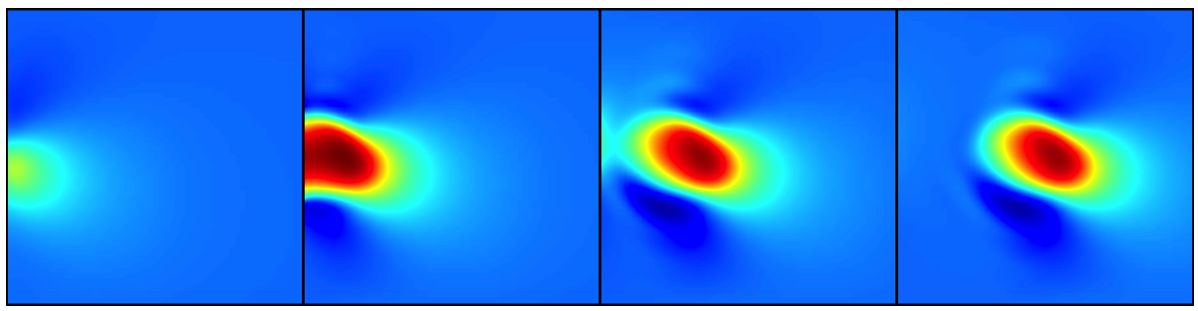}
         } \\
      a) \hspace{26.25mm} b)  \hspace{26.25mm} c) \hspace{23mm} d)
 \end{center}
  \caption{A single THV being generated at the inlet. Time is increasing with a constant time step from a) to d) and the inlet is on the left side of each frame.}
  \label{THV_time1}
\end{figure}

\subsection{Homogeneous Isotropic Turbulent Flow} \label{section_iso}
Large Eddy Simulation of isotropic turbulence was performed and the results were compared to experimental data collected by Comte-Bellot and Corrsin \cite{Comte_1,Comte_2}. Comte-Bellot and Corrsin measured the temporal decay of the turbulent kinetic energy of mesh-generated isotropic turbulence in a wind tunnel. For this case, a study of the spatial decay of the turbulent kinetic energy was used to validate the THV synthetic eddy method.

The numerical domain corresponds to the contracted test section of the wind tunnel downstream of the mesh. The dimensions of the domain were $120L_{mesh} \times 10L_{mesh} \times 10L_{mesh}$ in the streamwise, vertical, and spanwise directions, where $L_{mesh}$ is the size of the experimental mesh, which equals $0.0508$m  \cite{Comte_1}. In both the vertical and spanwise directions, the numerical domain size was chosen to be ten experimental mesh sizes with periodic boundary conditions, instead of using the entire cross-section of the wind tunnel test section. A uniform Cartesian grid consisting of $1200 \times 100 \times 100$ grid points was used. The increased artificial viscosity region was added for the last $20L_{mesh}$ in the streamwise direction at the outflow. A $M_{\infty} = 0.05$ uniform mean flow was imposed at the inflow, which is slightly larger than the experimental mean velocity ($12.7$ m/s); this is because the numerical algorithm is not able to handle very low Mach number flows. The non-dimensional streamwise turbulence intensity of the synthetic eddies imposed at the inlet corresponds to level reported at $42L_{mesh}$ in Comte-Bellot and Corrsin \cite{Comte_2}. 
The THV's were broken into ten generations, as given in Table \ref{THV_iso}, where the number of THV's in each generation was determined using equation (\ref{eddy_number}). The maximum radius of a THV was in the order of the experimental mesh size, $1.5L_{mesh}$, and the minimum radius was based on the grid resolution (five to six grid points across the smallest THV). Each THV generation is given an equal portion of the TKE, $b_m = 1/M$.
\begin{table}
 \begin{center}
  \begin{tabular}{|c|c|c|c|c|c|} \hline
       Generation  & Range of Radii & Number of THV's\\\hline
       $1$ 	&  $1.375\ L_{mesh} < a < 1.5\ L_{mesh}$	&  $8$  \\    \hline
       $2$ 	&  $1.25\ L_{mesh} < a < 1.375\ L_{mesh}$	&  $12$  \\    \hline
       $3$ 	&  $1.125\ L_{mesh} < a < 1.25\ L_{mesh}$	&  $22$  \\    \hline
       $4$ 	&  $1.0\ L_{mesh} < a < 1.125\ L_{mesh}$	&  $28$  \\    \hline
       $5$ 	&  $0.875\ L_{mesh} < a < 1.0\ L_{mesh}$	&  $36$  \\    \hline
       $6$ 	&  $0.75\ L_{mesh} < a < 0.875\ L_{mesh}$	&  $48$  \\    \hline
       $7$ 	&  $0.625\ L_{mesh} < a < 0.75\ L_{mesh}$	&  $68$  \\    \hline
       $8$ 	&  $0.5\ L_{mesh} < a < 0.625\ L_{mesh}$	&  $104$  \\    \hline
       $9$ 	&  $0.375\ L_{mesh} < a < 0.5\ L_{mesh}$	&  $180$  \\    \hline
       $10$	&  $0.25\ L_{mesh} < a < 0.375\ L_{mesh}$	&  $400$  \\    
       \hline
  \end{tabular}
  \caption{THV generations for the homogeneous cases}
  \label{THV_iso}
 \end{center}
\end{table}

Isosurfaces of Q-criterion are shown in Figure \ref{iso_Q_criterion1} . The THV's are generated at the inlet on the left and are convected downstream to the right. It can be seen from the isosurfaces that the shape of the generated THV's change from larger and compact structures near the inflow to smaller and stretched eddies farther downstream. In the proximity to the inflow boundary, one can hardly notice any transition from artificial to realistic turbulence. Comparing the regions near the inlet and outlet, there has been a noticeable dissipation of the eddies.

\begin{figure}
 \begin{center}
  \mbox{
      \includegraphics[width=16.cm, clip=true]{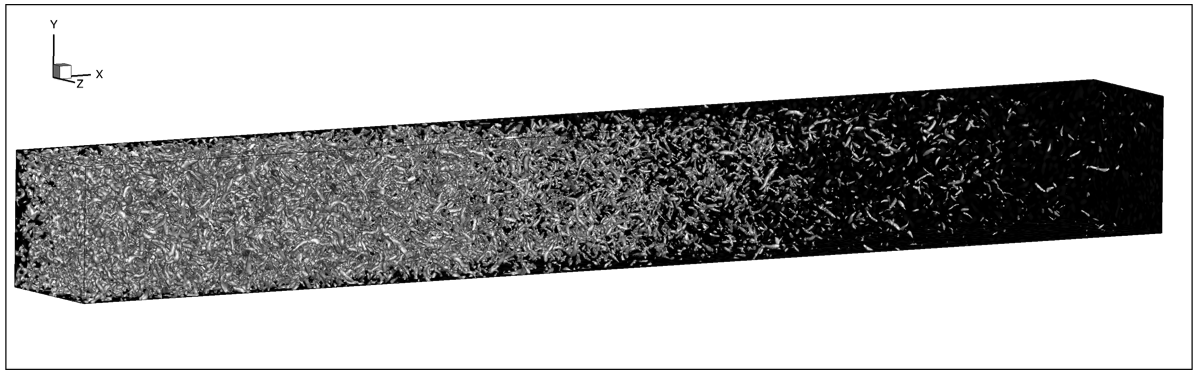}
         }  
 \end{center}
  \caption{Isosurfaces of Q-criterion for homogeneous isotropic turbulent flow (inlet is on the left).}
  \label{iso_Q_criterion1}
\end{figure}

Two-point spatial velocity correlations and non-dimensionalized turbulent kinetic energy spectrum of the synthetic fluctuations at the inlet plane are shown in Figure \ref{iso_inlet} along with experimental data collected at $42L_{mesh}$ by Comte-Bellot and Corrsin \cite{Comte_2}. Looking at the spatial correlations in Figure \ref{iso_inlet}(a), very good agreement is found for both the longitudinal and transverse correlations. Other than providing insight into what the maximum radius of the largest sized THV generation should be, the spatial correlations are not used as a control on the creation of the THV's. The energy spectrum in Figure \ref{iso_inlet}(b) also shows the same good agreement with the experimental data. Like with the spatial correlations, the only influence exerted over the energy spectrum of the created THV's is through the definition of the THV generations. While each THV generation is given an equal fraction of the total TKE, the wavenumber range inherently present in each THV generation is different. Because each THV is a coherent structure, each one contributes to the energy spectrum over the range of scales from the overall size of the THV to the smallest scale resolved by the grid.

\begin{figure}
 \begin{center}
      \includegraphics[width=6.8cm, clip=true]{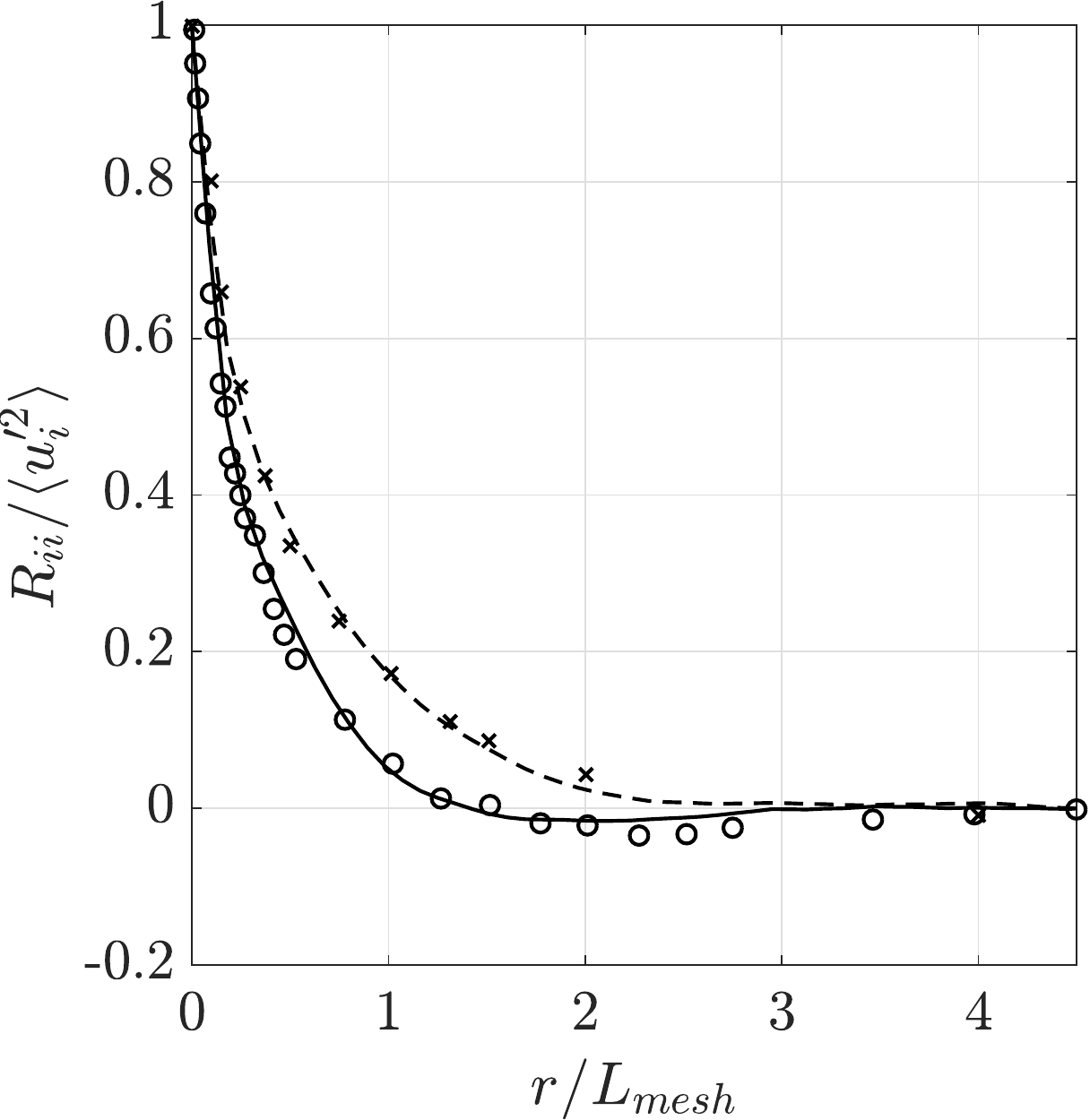}  \hspace{2mm}
      \includegraphics[width=7.0cm, clip=true]{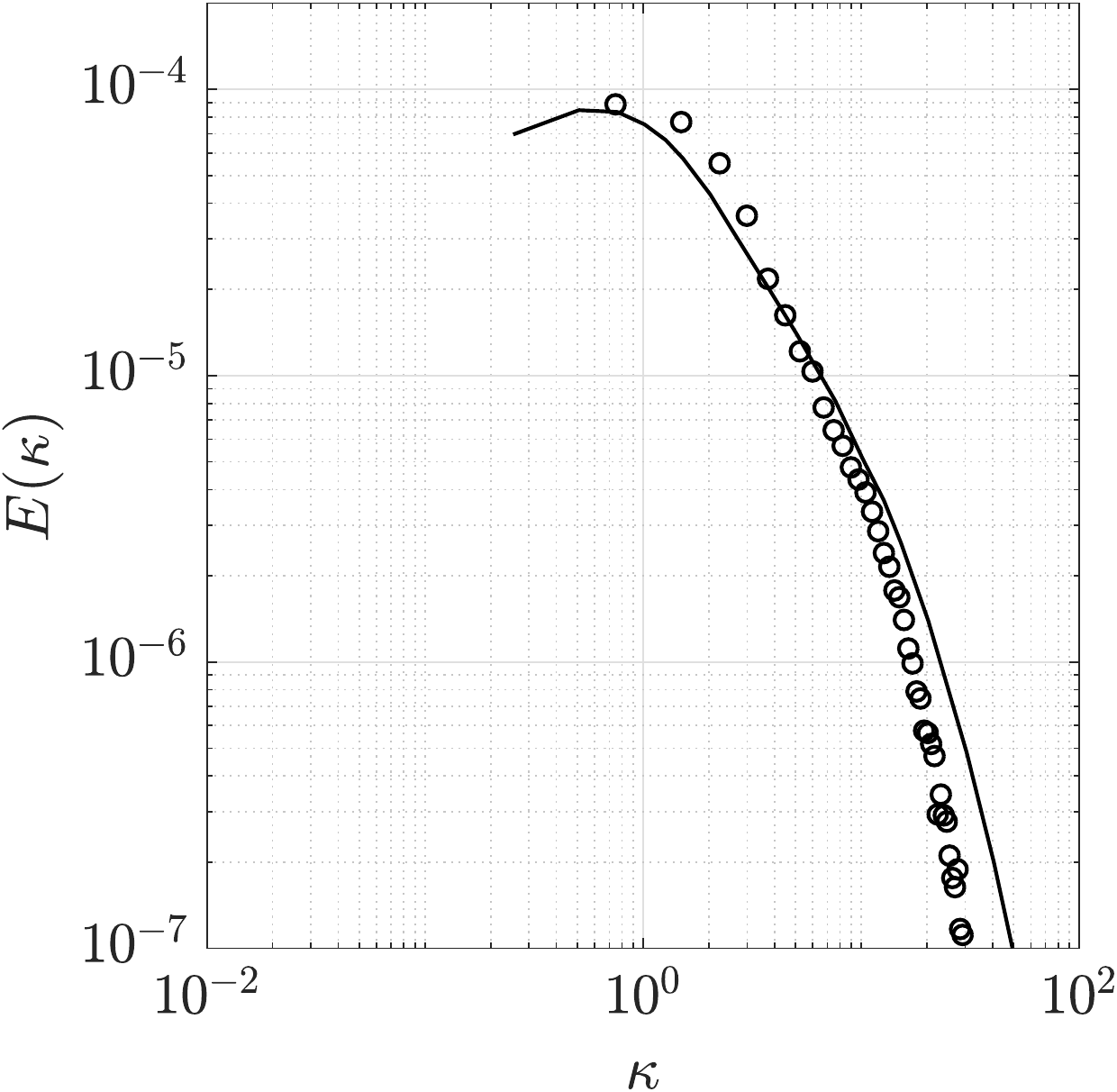}  \\
       a)  \hspace{56.25mm}  b)
 \end{center}
  \caption{Two-point spatial correlations (a) and turbulent kinetic energy spectrum as a function of wavenumber (b), captured at the inlet, compared with the experimental data collected by Comte-Bellot and Corrsin \cite{Comte_2}. a)\ -----) numerical transverse;\ -\ -\ -) numerical longitudinal;\ O) experimental transverse;\ X) experimental longitudinal. b) O)\ numerical; -----) experimental. } 
  \label{iso_inlet}
\end{figure}

Figure \ref{iso_TKE_skewness}(a) shows the spatial decay of the turbulent kinetic energy for the numerical results compared with the $k^{-1.25}$ power law fit to experimental data collected by Comte-Bellot and Corrsin \cite{Comte_1}. The THV SEM was able to reproduce the rate of turbulent kinetic energy decay seen in the experimental data. The lack of a recovery region just after the inlet further reinforces that the THV SEM is divergence-free. The slower decay of TKE present near the inlet is caused by the larger THV's at the inlet containing more of the TKE. This is seen in the energy spectrum at the inlet shown in Figure \ref{iso_inlet}(b), where the energy of the lower wavenumbers is slightly greater than the experimental results and the energy of the higher wavenumbers is slightly less than the experimental results.  Even though the filtering required by the numerical method for stability was keep to a minimum, this added dissipation contributed to the slightly quicker TKE decay.

The downstream development of the velocity derivative skewness is shown in Figure \ref{iso_TKE_skewness}(b) along with LES data from Jarrin et al. \cite{Jarrin_1} of spatially decaying homogeneous isotropic turbulence at a similar integral length scale Reynolds number. At the inlet, the skewness of the synthetic fluctuations generated using THV's is zero, which is expected because the THV's are symmetric. Moving shortly downstream, the skewness rapidly increases in magnitude for the flow generated by the THV SEM, while the skewness of the original SEM flow requires a much longer development length. Even though the Gaussian and tent function synthetic eddies of the original SEM and the THV's are all symmetric, the more physically based THV allows for the synthetic field to transition to realistic turbulence much quicker than the more abstract Gaussian shape.

\begin{figure}
 \begin{center}
      \includegraphics[width=7.0cm, clip=true]{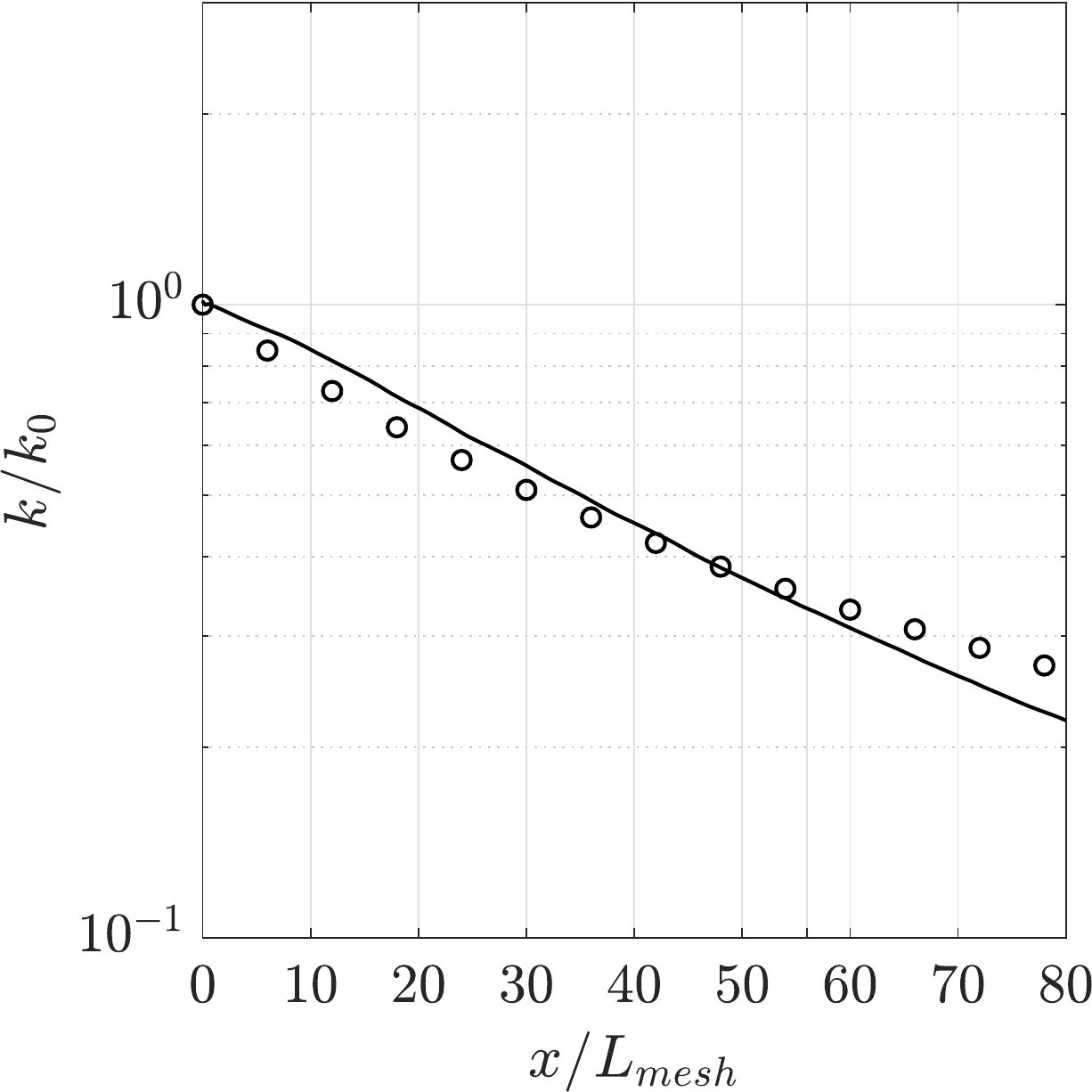}  \hspace{2mm}
      \includegraphics[width=6.8cm, clip=true]{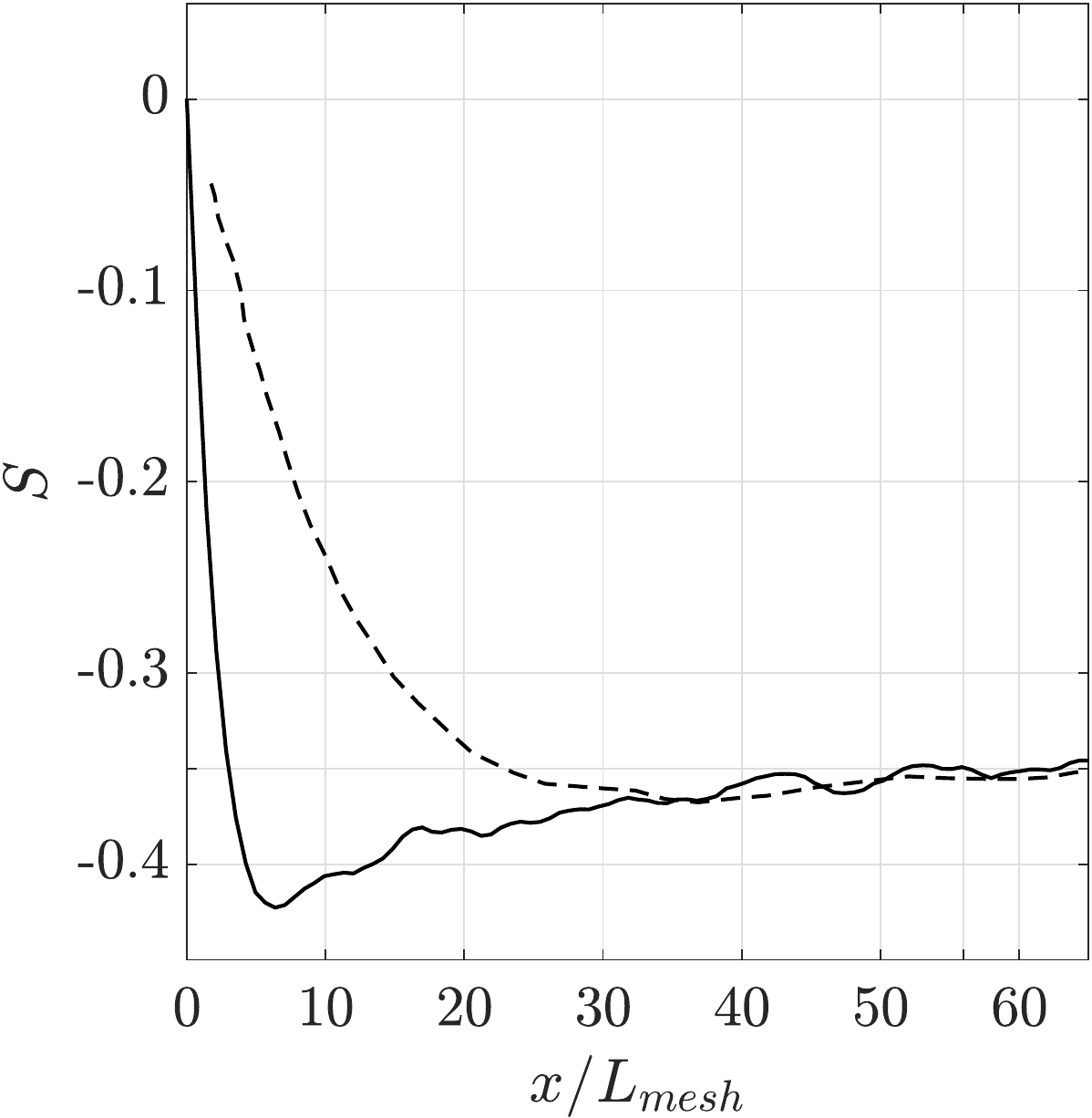}  \\
       a)  \hspace{56.25mm}  b)
 \end{center}
  \caption{Turbulent kinetic energy (a) and velocity derivative skewness (b) in the streamwise direction compared with the experimental data collected by Comte-Bellot and Corrsin \cite{Comte_1} and LES data from Jarrin et al. \cite{Jarrin_1} : -----) numerical; O) experimental;\ -\ -\ -) LES utilizing the original SEM.} 
  \label{iso_TKE_skewness}
\end{figure}

A comparison of the turbulent kinetic energy decay and the longitudinal two-point spatial correlation at the inlet (already presented in Figure \ref{iso_TKE_skewness}(a) and Figure \ref{iso_inlet}(a), respectively) with the LES data from Dietzel et al. \cite{Dietzel} is shown in Figure \ref{iso_compare_Dietzel}. Dietzel et al. \cite{Dietzel} simulated the temporal decay of homogeneous isotropic turbulence using the following three synthetic turbulence methods to generate synthetic initial conditions: the digital filtering method proposed by Klein et al. \cite{Klein} and improved by Kempf et al. \cite{Kempf_2}, the diffusion method of Kempf et al. \cite{Kempf_1}, and the Fourier method of Billson et al. \cite{Billson_1} and Davidson \cite{Davidson_2}. The decay of the TKE for the THV SEM exhibits similar behavior to the divergence-free (for isotropic turbulence) Fourier method, whereas the digital filtering and diffusion methods both show the significant dissipation immediately after synthetic fluctuations are imposed that is characteristic of methods that do not satisfy the divergence-free condition.

\begin{figure}
 \begin{center}
      \includegraphics[width=6.8cm, clip=true]{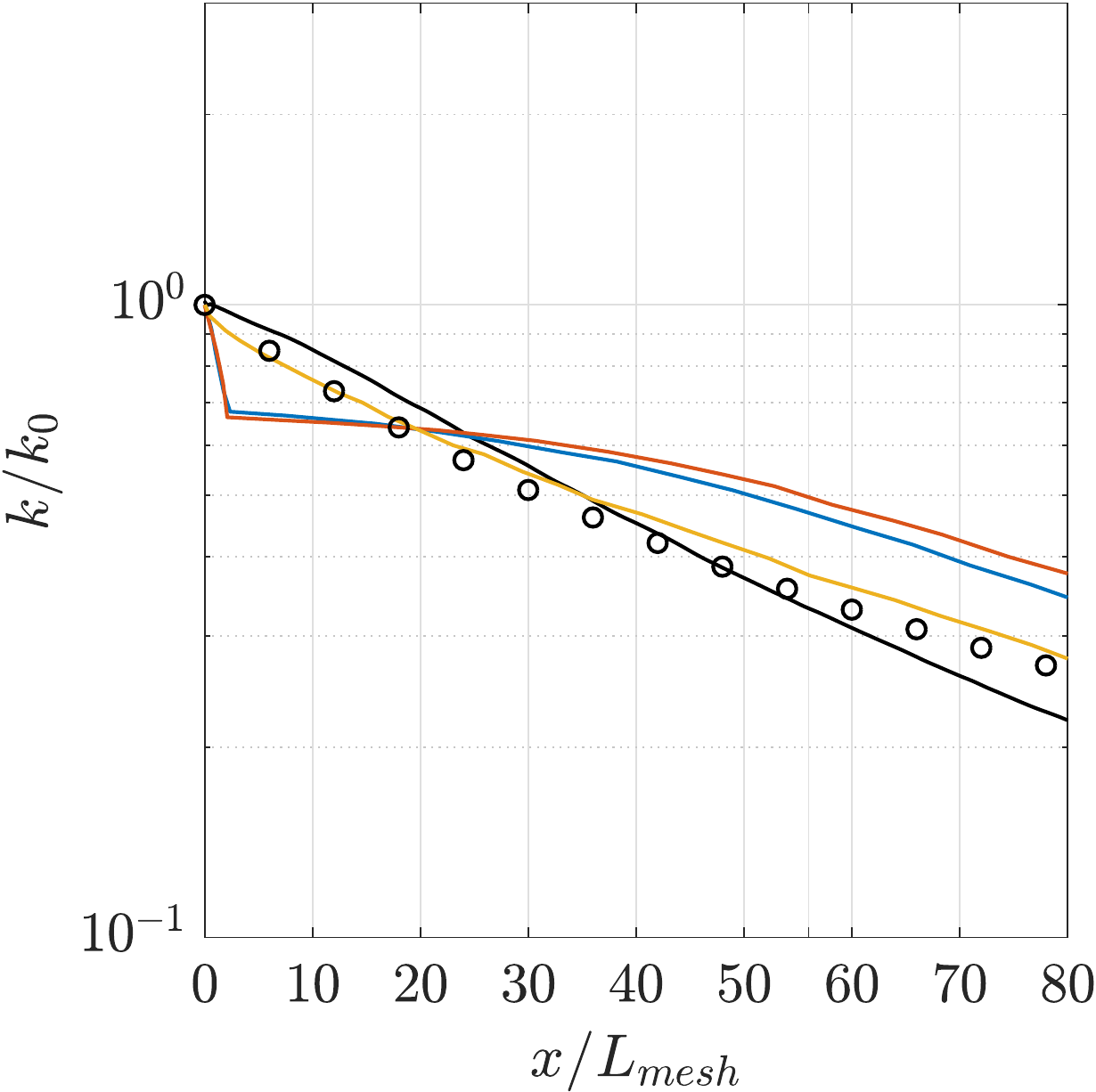}  \hspace{2mm}
      \includegraphics[width=6.8cm, clip=true]{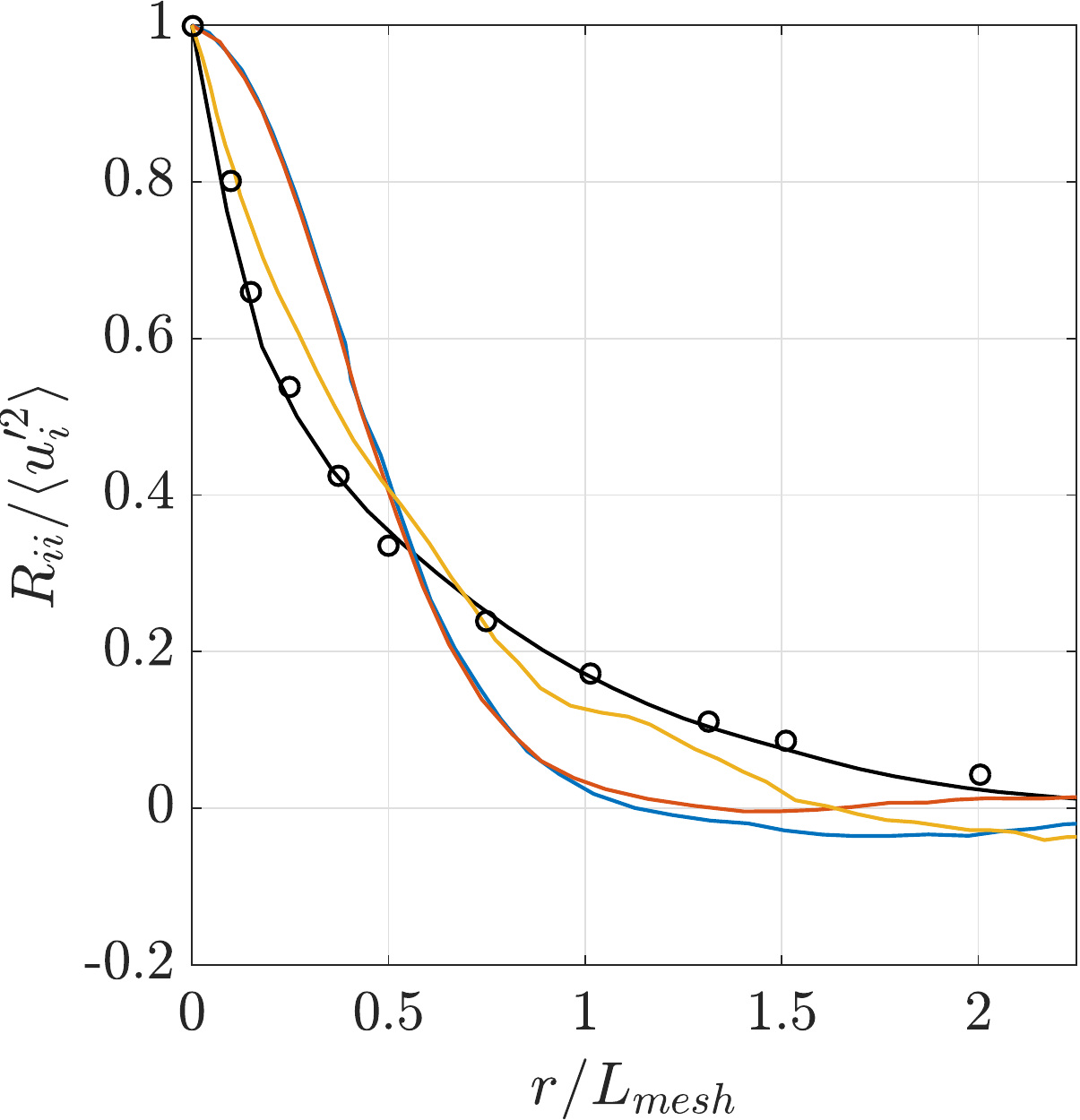}  \\
       a)  \hspace{56.25mm}  b)
 \end{center}
  \caption{Turbulent kinetic energy in the streamwise direction (a) and longitudinal two-point spatial correlation at the inlet compared with the experimental data collected by Comte-Bellot and Corrsin \cite{Comte_1} and LES data from Dietzel et al. \cite{Dietzel}: black) THV SEM; O) experimental;\ blue) digital filtering method \cite{Klein,Kempf_2};\ red) diffusion method \cite{Kempf_1}; yellow) Fourier method \cite{Billson_1,Davidson_2}.} 
  \label{iso_compare_Dietzel}
\end{figure}

Two-point spatial velocity correlations and non-dimensionalized turbulent kinetic energy spectrum $56L_{mesh}$ downstream of the inlet plane are shown in Figure \ref{iso_98} along with experimental data collected at the corresponding downstream location by Comte-Bellot and Corrsin \cite{Comte_2}. Again, very good agreement with the experimental results is seen for both the spatial correlations in Figure \ref{iso_98}(a) and the energy spectrum in Figure \ref{iso_98}(b). The well-modeled synthetic fluctuations imposed by the THV SEM at the inlet quickly developed into realistic turbulence and correctly reproduced downstream statistics.

\begin{figure}
 \begin{center}
      \includegraphics[width=6.8cm, clip=true]{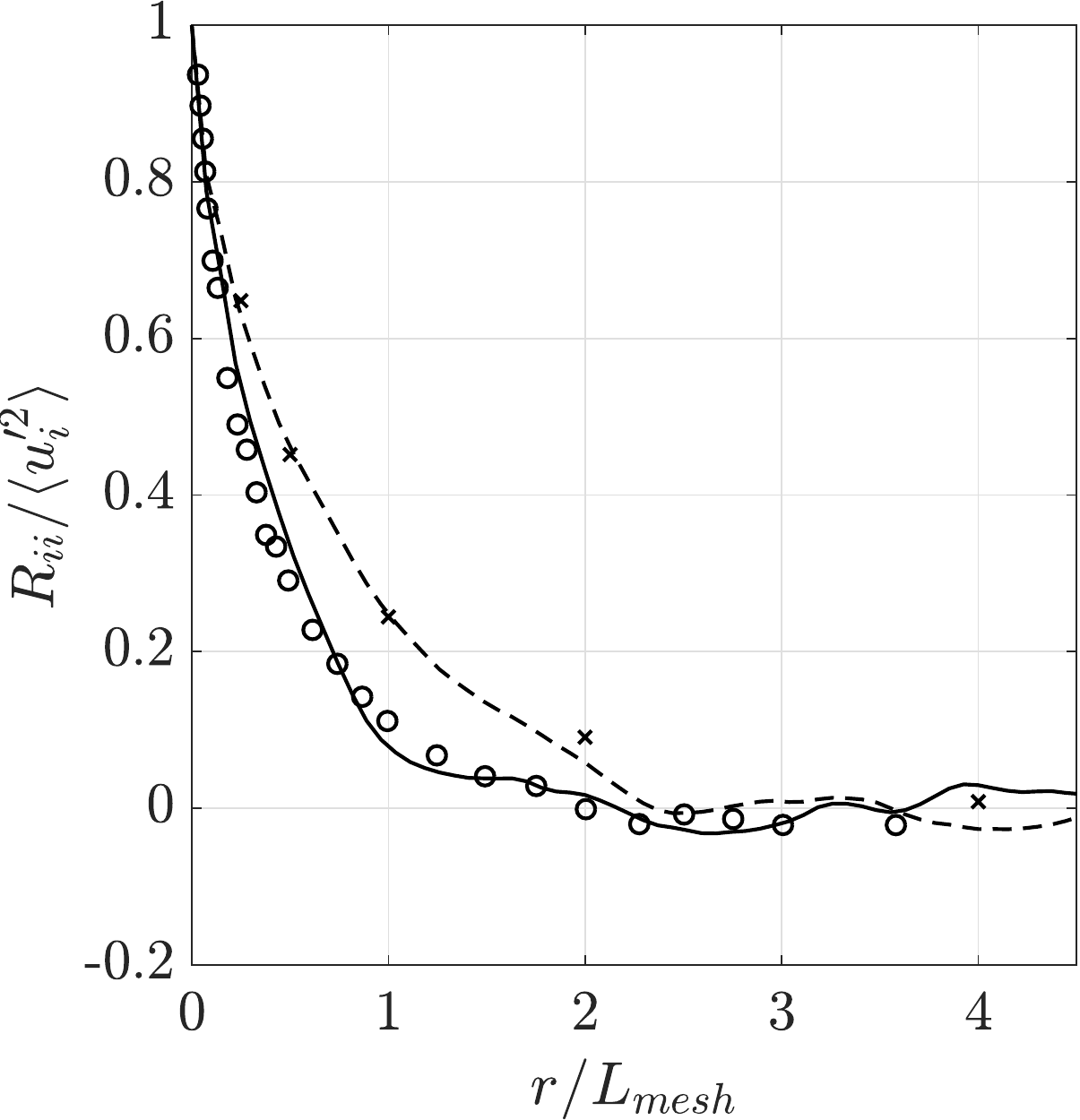}  \hspace{2mm}
      \includegraphics[width=7.0cm, clip=true]{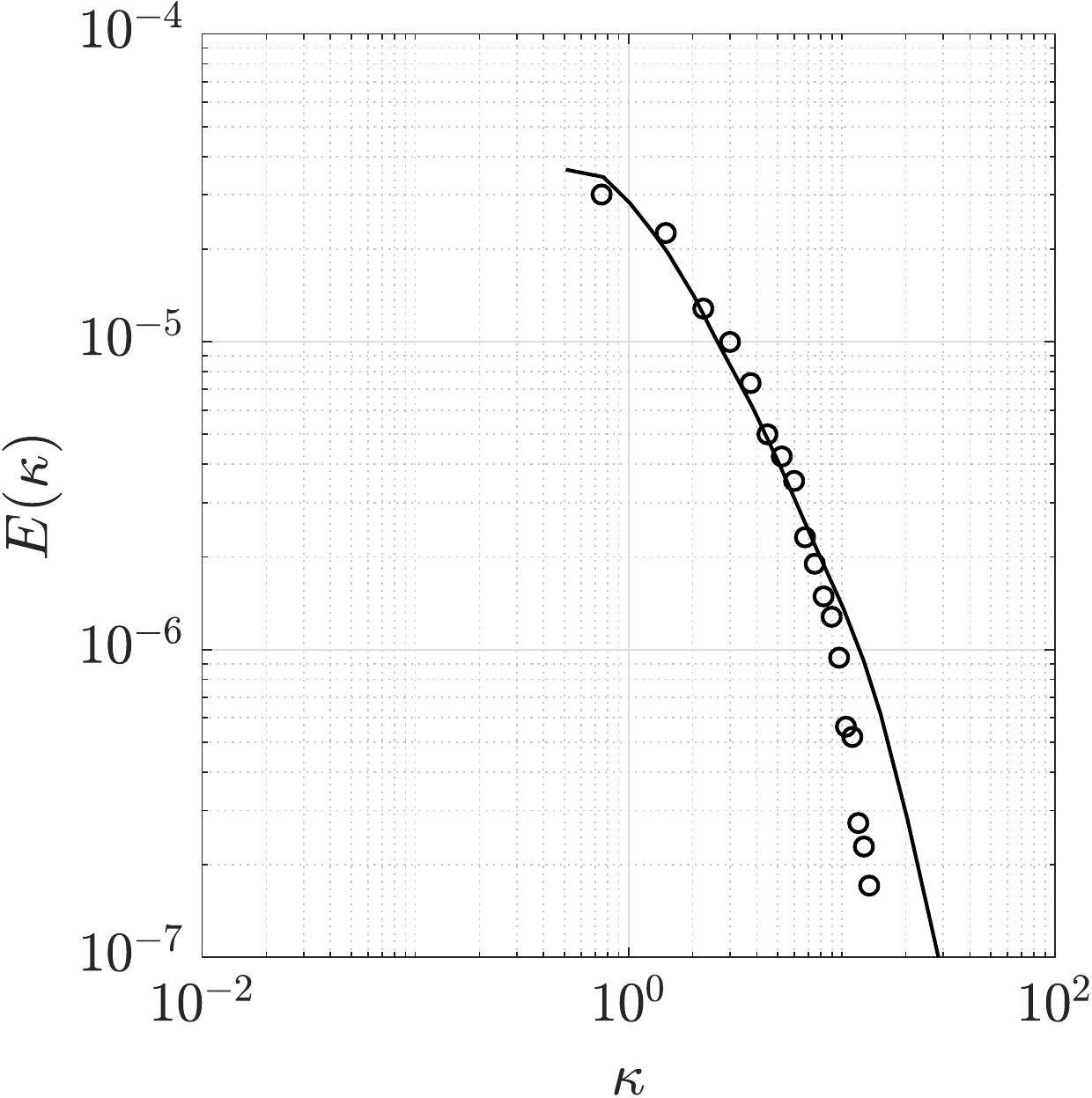}  \\
       a)  \hspace{56.25mm}  b)
 \end{center}
  \caption{Two-point spatial correlations (a) and turbulent kinetic energy spectrum as a function of wavenumber (b), captured $56L_{mesh}$ downstream of the inlet, compared with the experimental data collected by Comte-Bellot and Corrsin \cite{Comte_2}. a)\ -----) transverse;\ -\ -\ -) longitudinal;\ O) experimental transverse;\ X) experimental longitudinal. b) O)\ numerical; -----) experimental.} 
  \label{iso_98}
\end{figure}

\subsection{Turbulent Channel Flow}
To further test the capability of the THV SEM to model anisotropic non-homogeneous turbulence, a turbulent channel flow case was considered. Reynolds stress tensor mean velocity profiles from the DNS of a fully turbulent channel at $Re_{\tau}=395$ collected by Moser et al. \cite{Moser} were imposed at the inlet plane. The dimensions of the domain were $15\delta \times 2\delta \times 3\delta$ in streamwise, vertical, and spanwise directions, where $\delta$ is the channel half height. A Cartesian grid composed of $200 \times 100 \times 100$ grid points was used. The grid was uniformly spaced in the streamwise and spanwise directions, and stretched in the vertical direction in order to increase resolution at the walls. This allowed for grid spacings of $\Delta x^+ = 30$, $\Delta z^+ = 12$, $\Delta y_{min}^+ = 1$, and $\Delta y_{max}^+ = 17.5$. No-slip wall boundary conditions were used on the bottom and top boundaries, and periodic boundary conditions were applied in the spanwise direction. The increased artificial viscosity region was added for the last $3\delta$ in the streamwise direction at the outflow. $6218$ THV's were imposed at any one instant in time on to the mean flow at the inflow boundary. The maximum radius allowed was $0.6\delta$, while the minimum radius was $0.006\delta$. The THV's were split into five generations, as given in Table \ref{THV_channel}, with the first generation being the elongated eddies described in Section \ref{section_stretch} and the fifth generation being clustered very near the walls. This clustering of THV's replicates the smallest turbulent structures very near the wall. Because the energy spectra are not global quantities and a relationship between the factor $b_m$ and distance from the wall was not readily apparent, $b_m = 1$.

\begin{table}
 \begin{center}
  \begin{tabular}{|c|c|c|c|c|c|} \hline
       Generation  & Range of Radii & Range of y & Number of THV's\\\hline
       $1$  &  $0.24\ \delta < a < 0.32\ \delta$     	&  $0.64\ \delta < |y| < 0.68\ \delta$ 		&  $30$  \\    \hline
       $2$  &  $0.4\ \delta < a < 0.6\ \delta$  	&  $0 < |y| < (\delta - a)$ 		&  $8$  \\    \hline
       $3$  &  $0.1\ \delta < a < 0.2\ \delta$   	&  $0 < |y| < (\delta - a)$ 		&  $180$  \\    \hline
       $4$  &  $0.02\ \delta < a < 0.1\ \delta$ 	&  $0 < |y| < (\delta - a)$ 		&  $4000$  \\    \hline
       $5$  &  $0.006\ \delta < a < 0.01\ \delta$   	&  $0.9\ \delta < |y| < 0.99\ \delta$ 	&  $2000$  \\     
       \hline
  \end{tabular}
  \caption{THV generations for the channel case, where $-\delta < y < \delta$.}
  \label{THV_channel}
 \end{center}
\end{table}

Contours of streamwise velocity and vorticity magnitude at the inflow plane are shown in Figure \ref{channel_u_vort_mag}. These contours show the synthetic inflow, the superposed distorted THV's, that is imposed by the THV SEM. Notice how the concept of the generations of THV's manifests itself in the synthetic inflow with a few large eddies surrounded by ever smaller structures. Moving from the center of the channel to the walls, the size the imposed eddies decreases. The vorticity magnitude contour especially illuminates the effect of the clustered eddies near the wall, where the vorticity magnitude is larger as compared to the center of the channel. The development of the elongated turbulent structures stretching downstream from the walls can be seen from the Q-criterion isosurfaces in Figure \ref{channel_Q_criterion}. While the synthetic eddies are released into the the domain and quickly evolve into realistic turbulent structures, there is still a small region near the inlet where the non-stretched THV's of generations two through five need to develop.

\begin{figure}
 \begin{center}
      \includegraphics[width=6.8cm, clip=true]{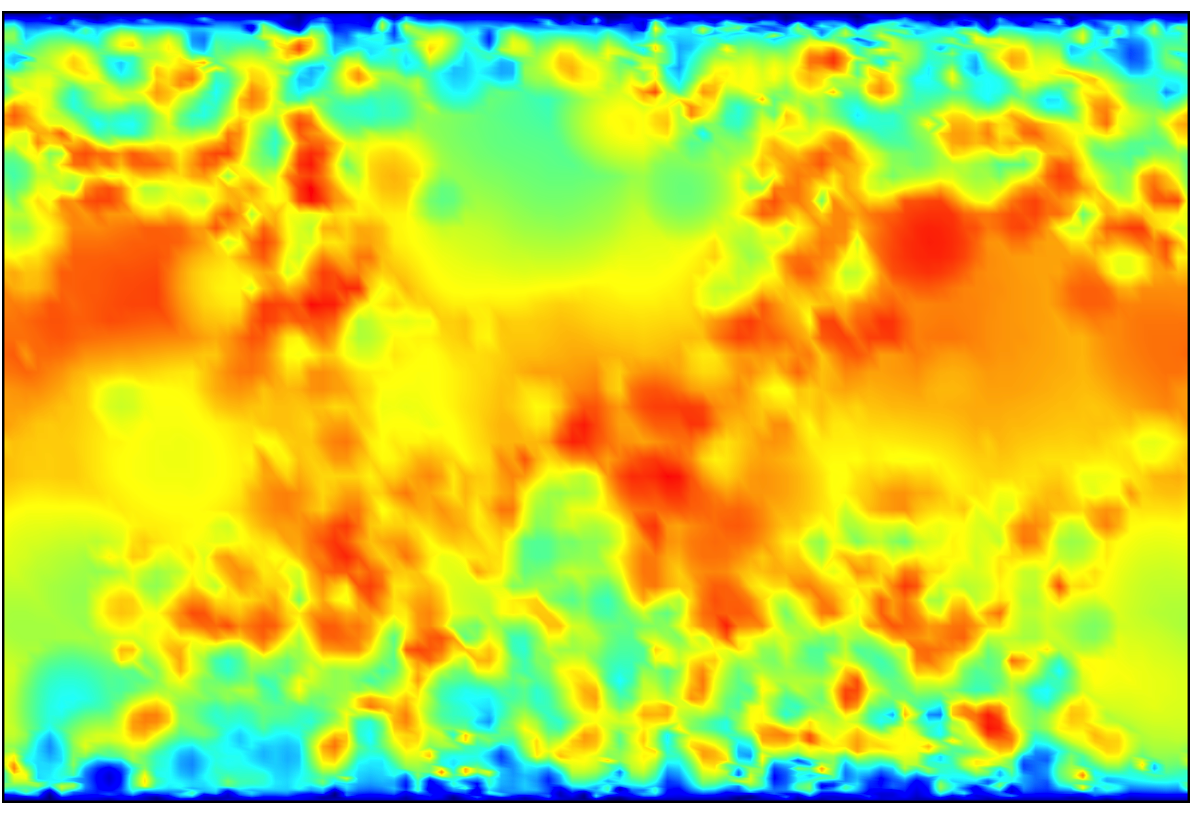}  \hspace{2mm}
      \includegraphics[width=6.8cm, clip=true]{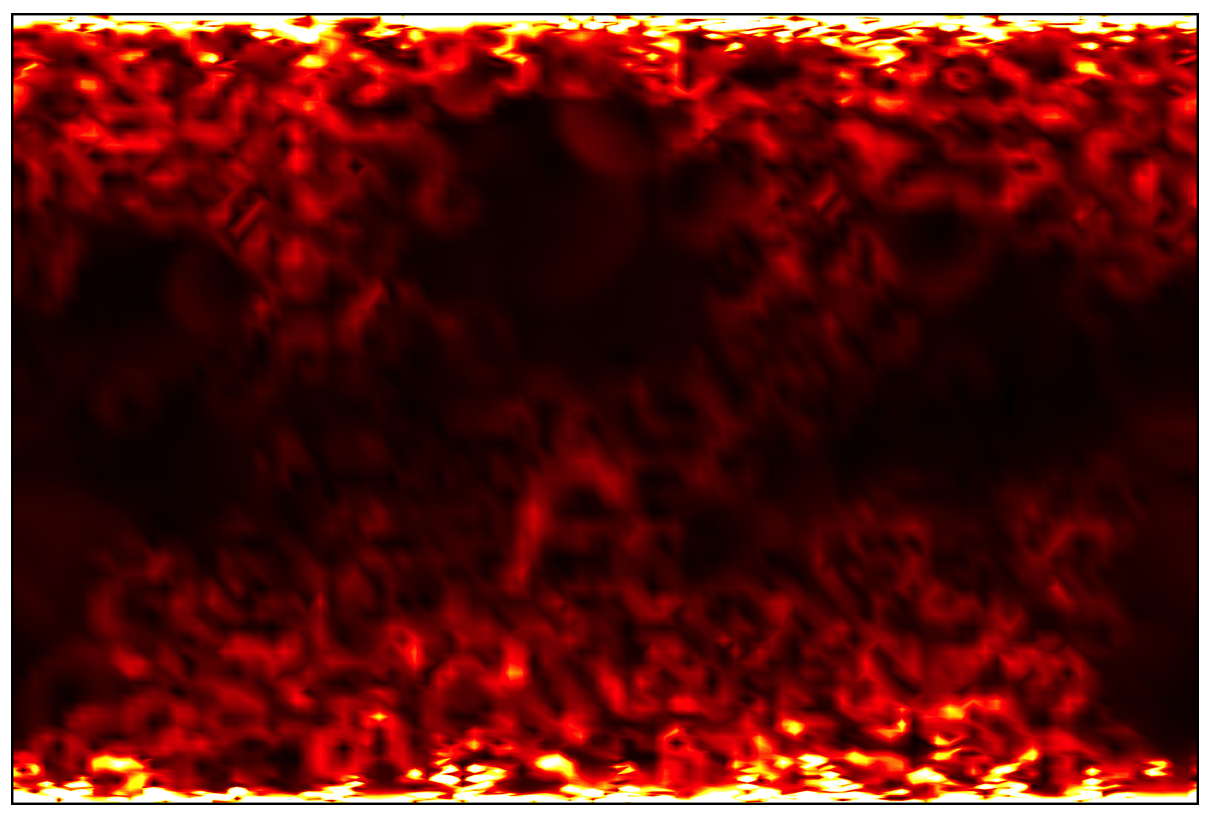}  \\
       a)  \hspace{56.25mm}  b)
 \end{center}
  \caption{Contours for turbulent channel flow at the inlet plane(from the synthetic turbulence model): a) streamwise velocity; b) vorticity magnitude.} 
  \label{channel_u_vort_mag}
\end{figure}

\begin{figure}
 \begin{center}
  \mbox{
      \includegraphics[width=15.cm, clip=true]{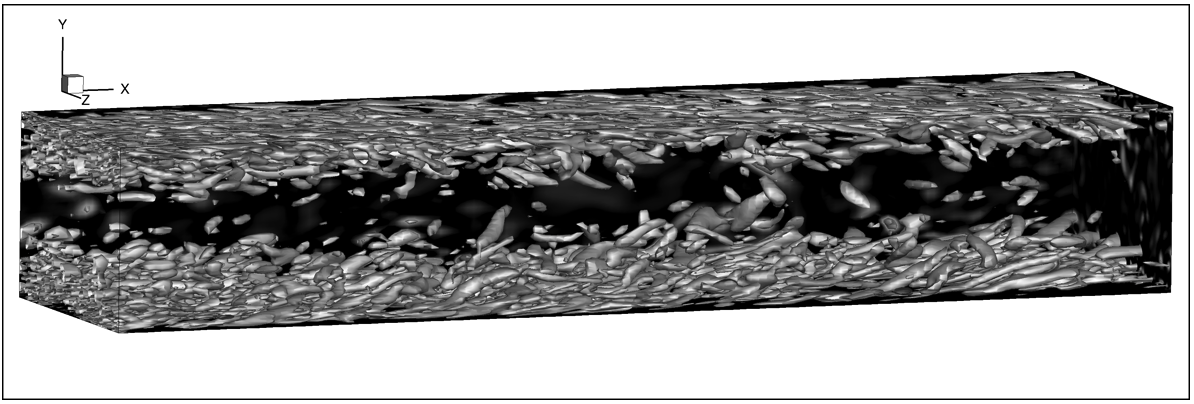}
         }  
 \end{center}
  \caption{Isosurfaces of Q-criterion for turbulent channel flow.}
  \label{channel_Q_criterion}
\end{figure}

Vertical profiles of the mean streamwise velocity and Reynolds stresses are plotted in Figure \ref{channel_u_Re_stress_inlet}, where averaging in both time and the homogeneous spanwise directions has been taken. As seen in Figure \ref{channel_u_Re_stress_inlet}(a), the DNS mean streamwise velocity profile is well reproduced. When looking at the Reynolds stresses in Figure \ref{channel_u_Re_stress_inlet}(b), there is excellent agreement between the heights of $-0.9\delta$ and $0.9\delta$. This height range corresponds to the effective region where most of the THV's are imposed. A portion of the smaller THV's of the fourth generation and the fifth clustered generation are imposed at heights closer to the wall, where their effect is clearly seen in the bulge of the $\langle u'u' \rangle$ profile very close to the walls, but those THV's are not enough to reproduce the DNS profiles. With more smaller and smaller THV's in those nearest wall regions, those Reynolds stress profiles will be better matched.

\begin{figure}
 \begin{center}
      \includegraphics[width=6.8cm, clip=true]{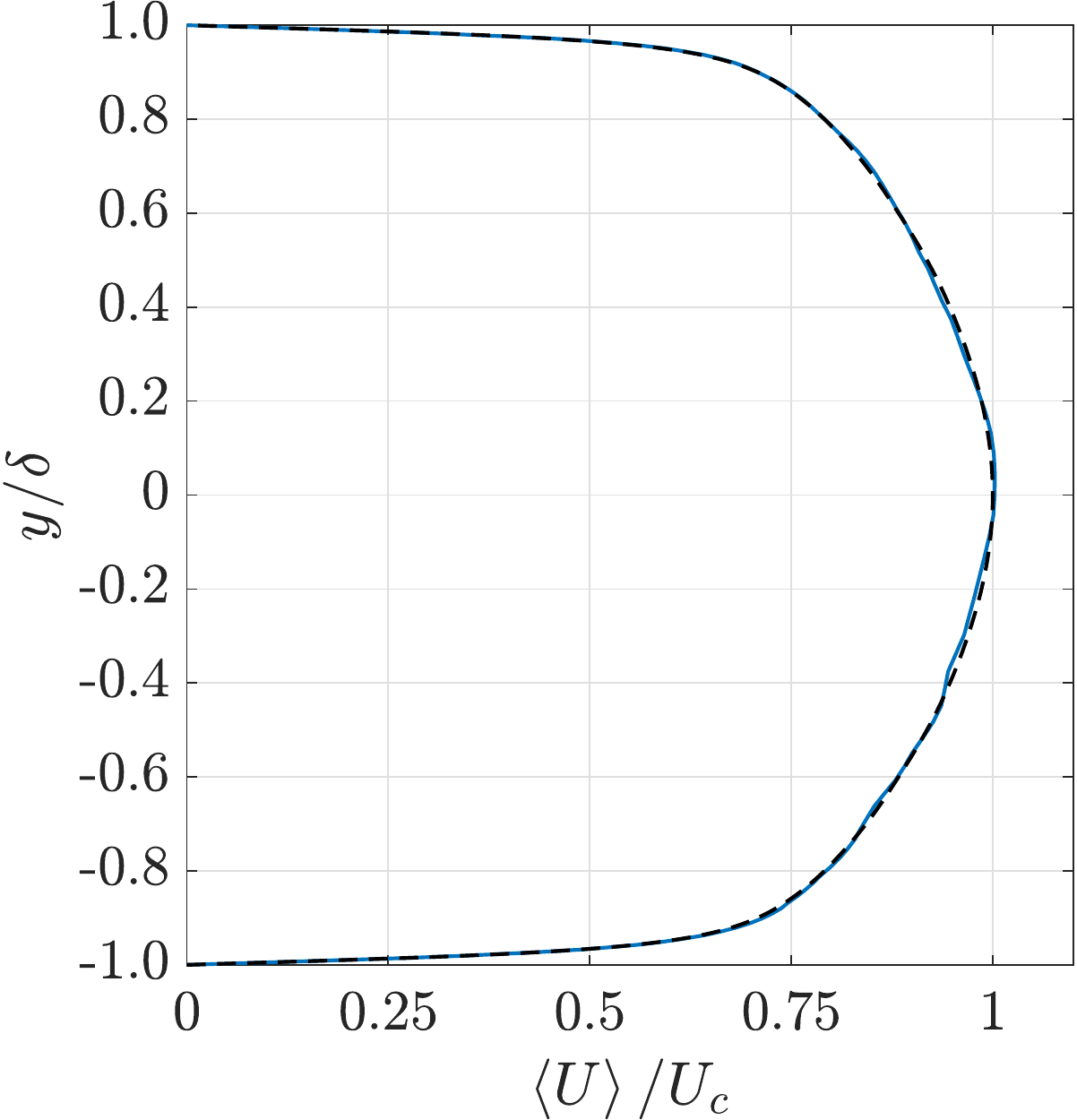}  \hspace{2mm}
      \includegraphics[width=6.8cm, clip=true]{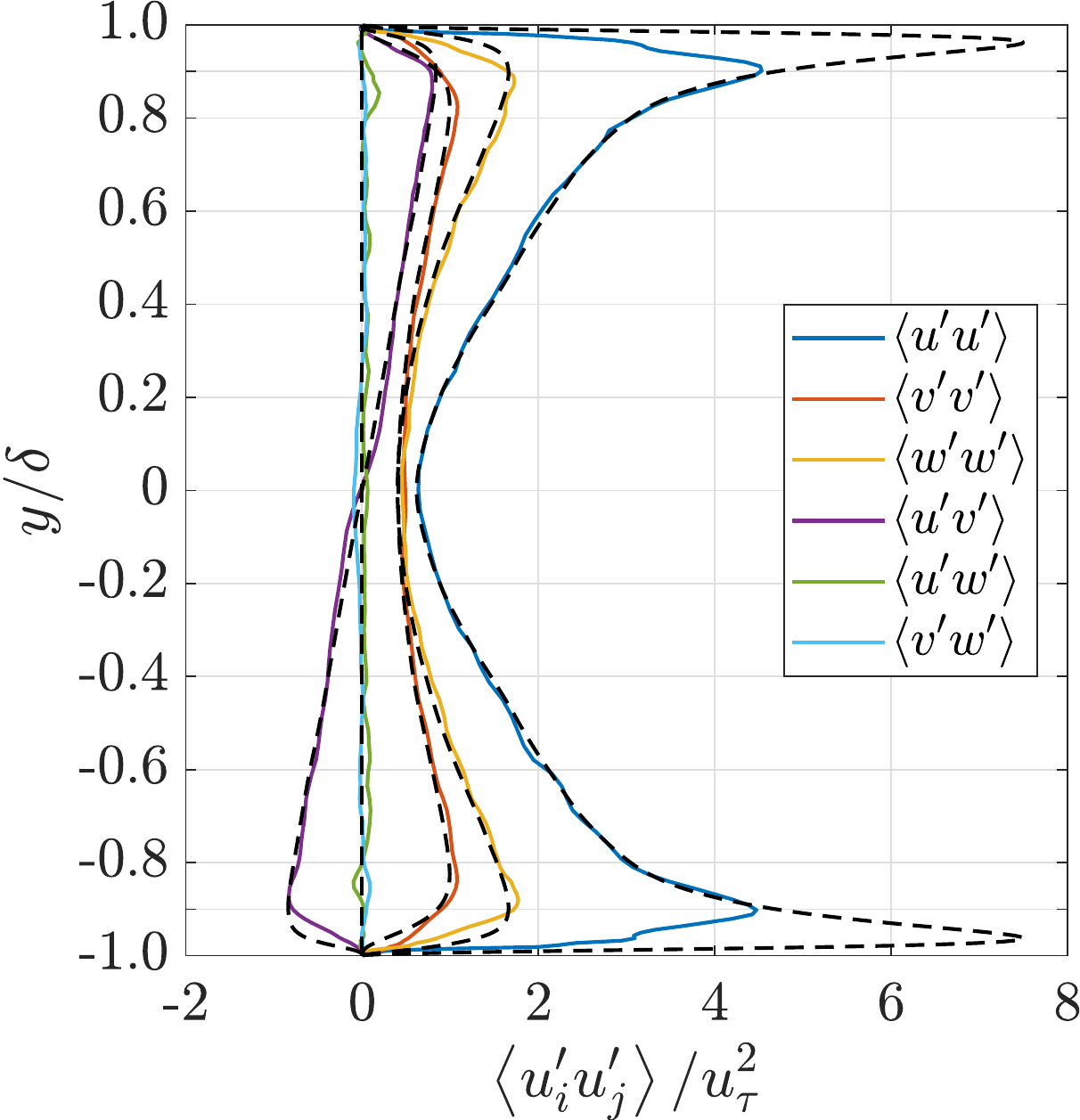}  \\
       a)  \hspace{56.25mm}  b)
 \end{center}
  \caption{Vertical profiles at the inlet compared with the DNS data collected by Moser et al. \cite{Moser} of: a) mean streamwise velocity; b) Reynolds stresses;\ -----) numerical;\ -\ -\ -) DNS data.} 
  \label{channel_u_Re_stress_inlet}
\end{figure}

Two-point spanwise spatial velocity correlations of the synthetic fluctuations at the inlet plane and at three different heights from the wall are shown in Figure \ref{channel_Rij_inlet} along with the corresponding DNS data from Moser et al. \cite{Moser}. There is generally good agreement at spanwise distances less than $0.4\delta$. Closer to the wall, Figure \ref{channel_Rij_inlet}(a), the synthetic fluctuations are composed of the smaller THV's. They are only locally correlated, so the spatial correlations quickly go to zero. Moving away from the wall, Figure \ref{channel_Rij_inlet}(b), and towards the center of the channel, Figure \ref{channel_Rij_inlet}(c), the largest THV's become more present. This is seen in the higher correlations at larger spanwise distances, which might be a hint that the largest THV's are too large in size. As a general trend across all three heights from the wall, the THV's do not reproduce negative spatial correlations, especially at larger spanwise distances. The transverse correlation $0.2\delta$ from the wall does dip slightly below zero, which was also seen for the transverse correlation at the inlet of the homogeneous isotropic turbulence (Figure \ref{iso_inlet}(a)), but it does not reproduce the negative correlation.

\begin{figure}
 \begin{center}
     \includegraphics[width=5.0cm, clip=true]{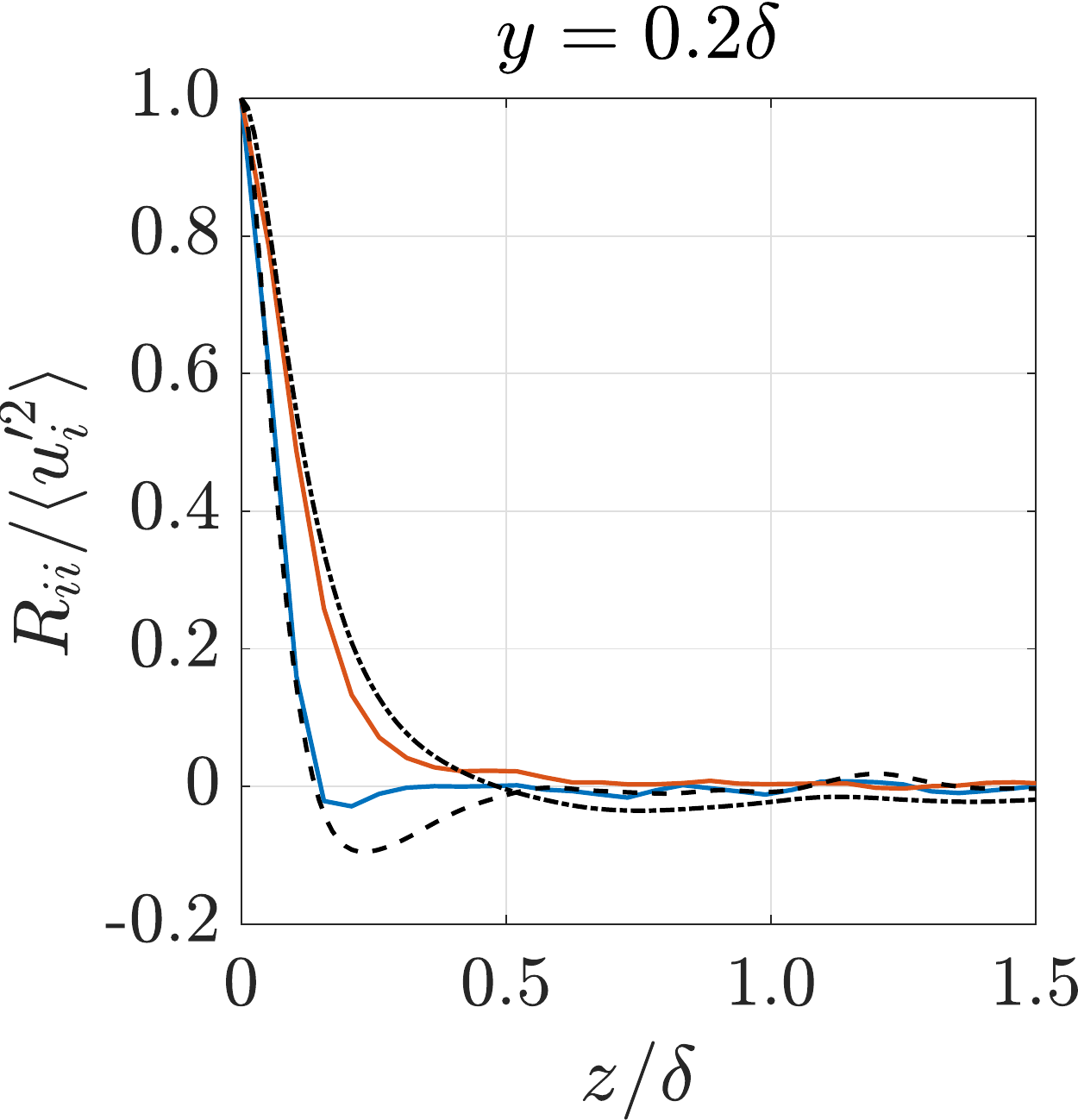}  \hspace{2mm}
     \includegraphics[width=5.0cm, clip=true]{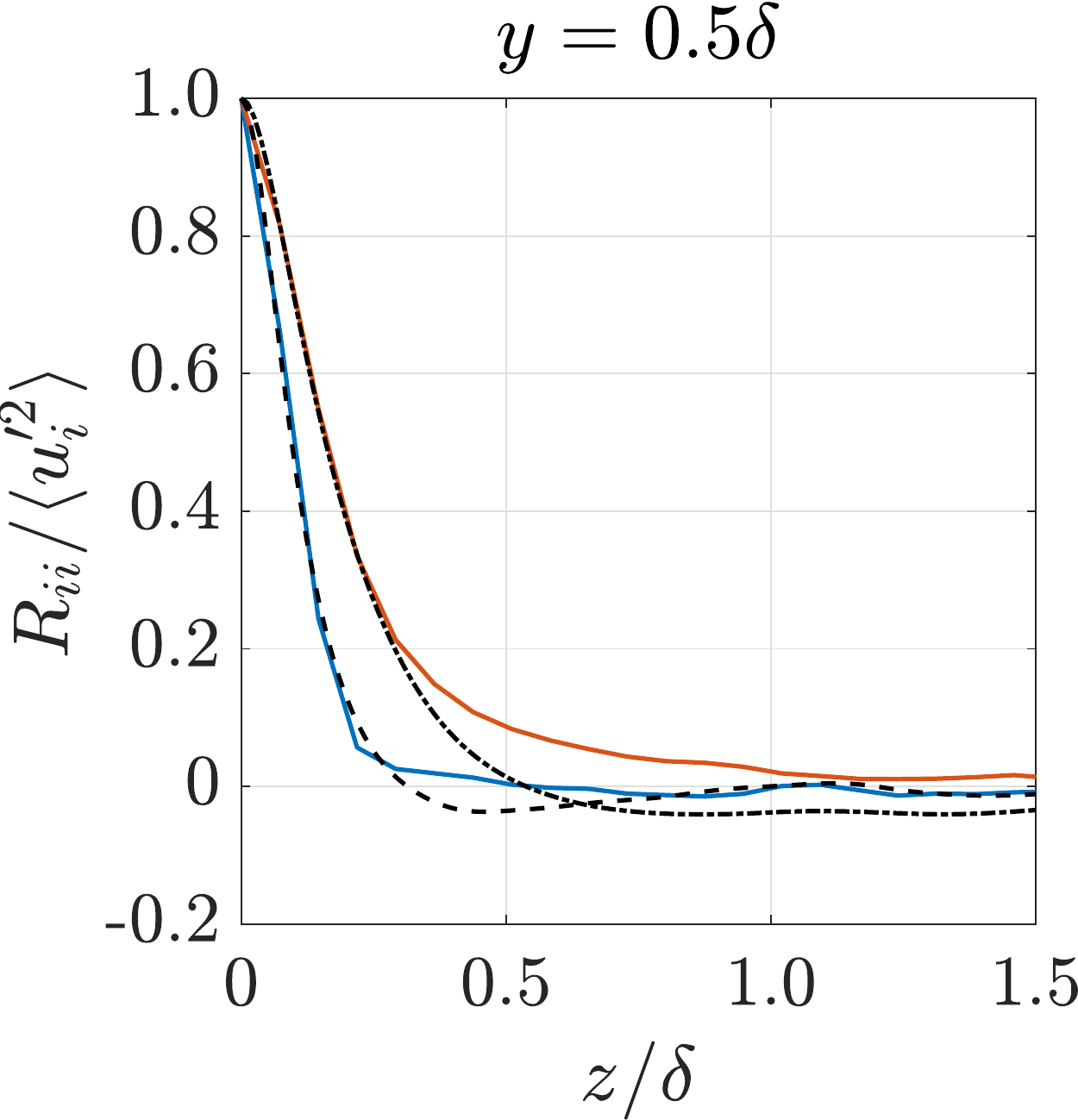}  \hspace{2mm}
     \includegraphics[width=5.0cm, clip=true]{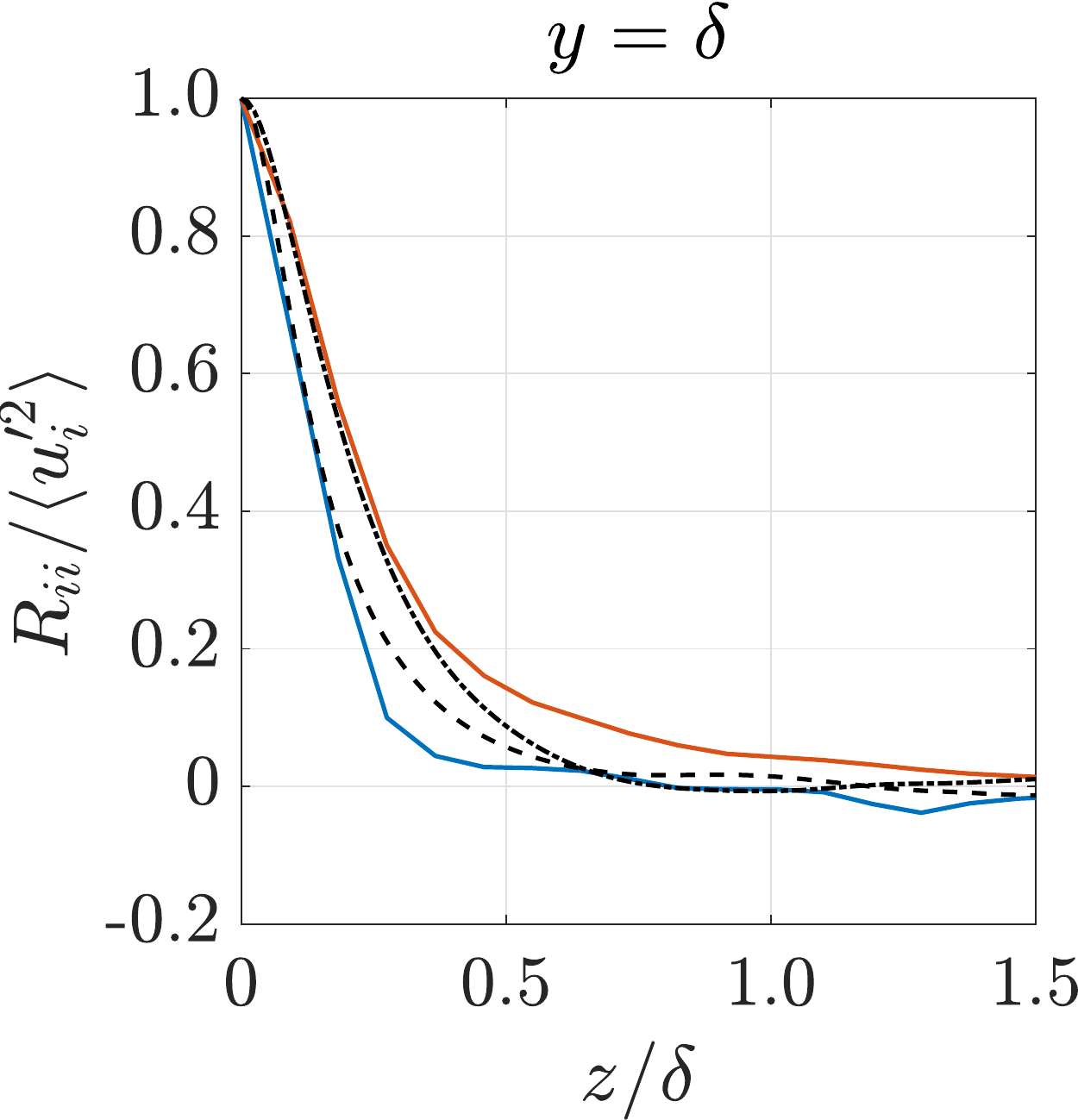}  \\
      \hspace{1.25mm} a)  \hspace{47.mm}  b)  \hspace{47.mm} c)
 \end{center}
  \caption{Two-point spanwise spatial correlations, captured at the inlet and at three different heights from the wall , compared with the DNS data collected by Moser et al. \cite{Moser}. \ blue) transverse;\ red) longitudinal;\ -\ -\ -) DNS transverse;\ -.-.-) DNS longitudinal.}
  \label{channel_Rij_inlet}
\end{figure}

Figure \ref{channel_E_inlet} shows the non-dimensionalized spanwise turbulent kinetic energy spectra of the synthetic fluctuations at the inlet plane and at three different heights from the wall along with the corresponding DNS data from Moser et al. \cite{Moser}. There is good agreement at all three heights in the higher wavenumbers, with better agreement when moving towards the wall. Unlike for homogeneous turbulence where each THV generation matches an equal fraction of the TKE at any point, because of the non-homogeneous nature of the wall normal direction, each individual THV matches the total TKE at any point. So, as seen in the darker large circular areas in the inlet vorticity contour of Figure \ref{channel_u_vort_mag}(b), the generations of smaller THV's are not created inside as many of the largest THV's which does not allow for the same sort of fine tuning of the energy spectrum as for the homogeneous turbulence. This causes the general under-prediction of the peaks of the energy spectra. Looking back again to Figure \ref{channel_u_vort_mag}(b) and considering the larger THV present towards the middle top of the contour, these larger THV's match the Reynolds stresses away from the walls, but can contribute disproportionately to energy at the lowest wavenumbers. This is why the spectra of the synthetic fluctuations levels off at the lowest wavenumbers.

\begin{figure}
 \begin{center}
     \includegraphics[width=5.0cm, clip=true]{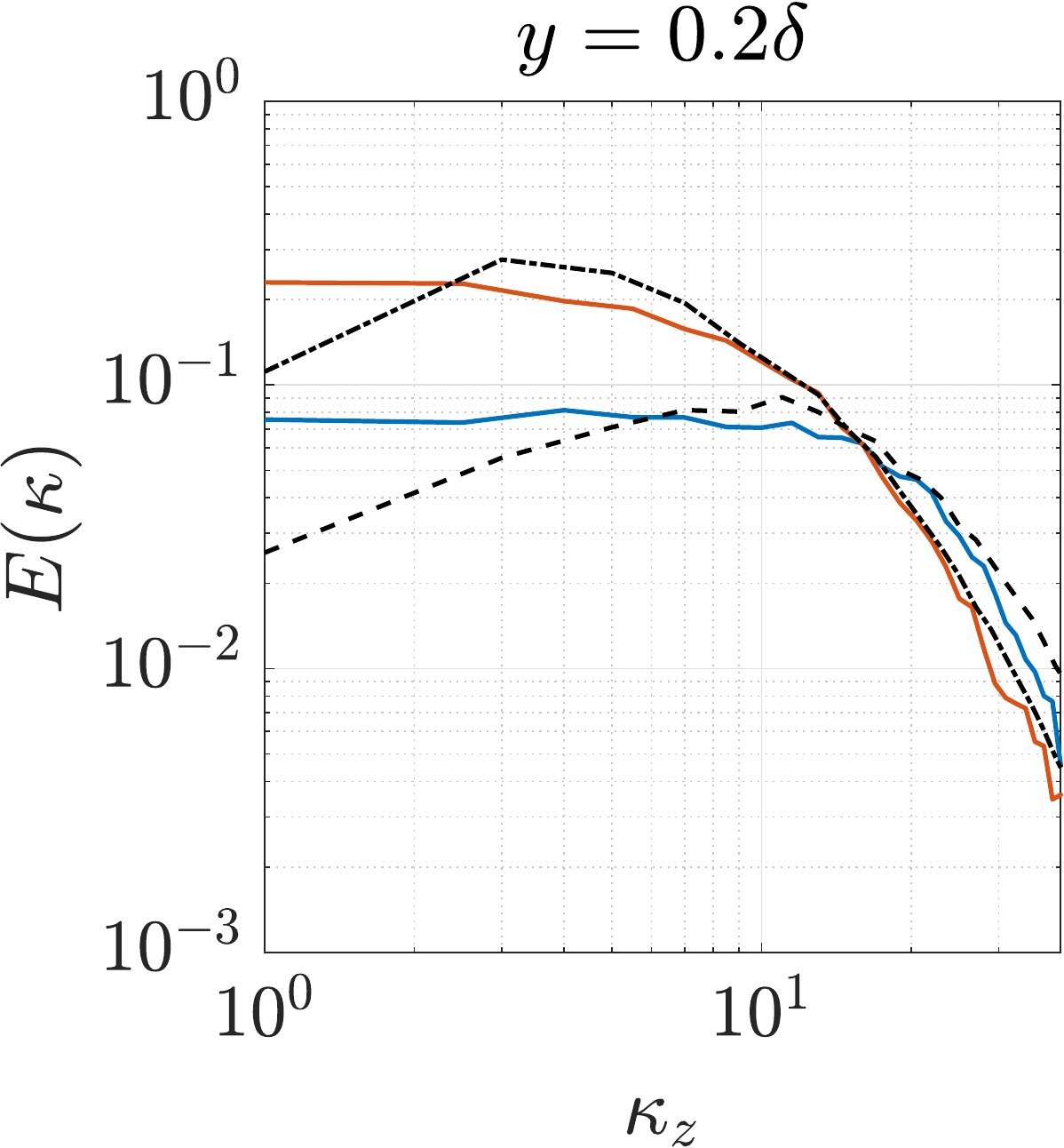}  \hspace{2mm}
     \includegraphics[width=5.0cm, clip=true]{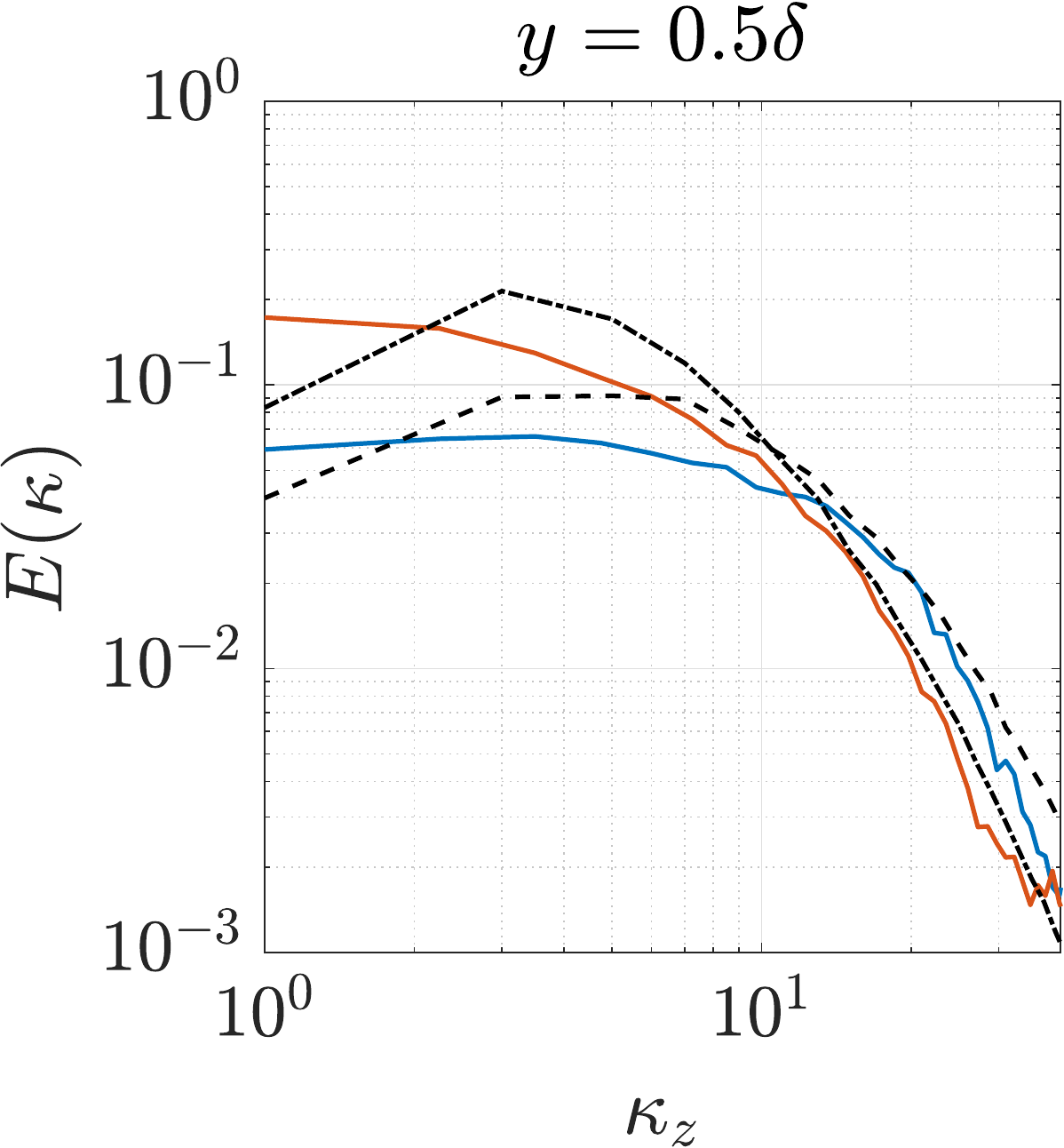}  \hspace{2mm}
     \includegraphics[width=5.0cm, clip=true]{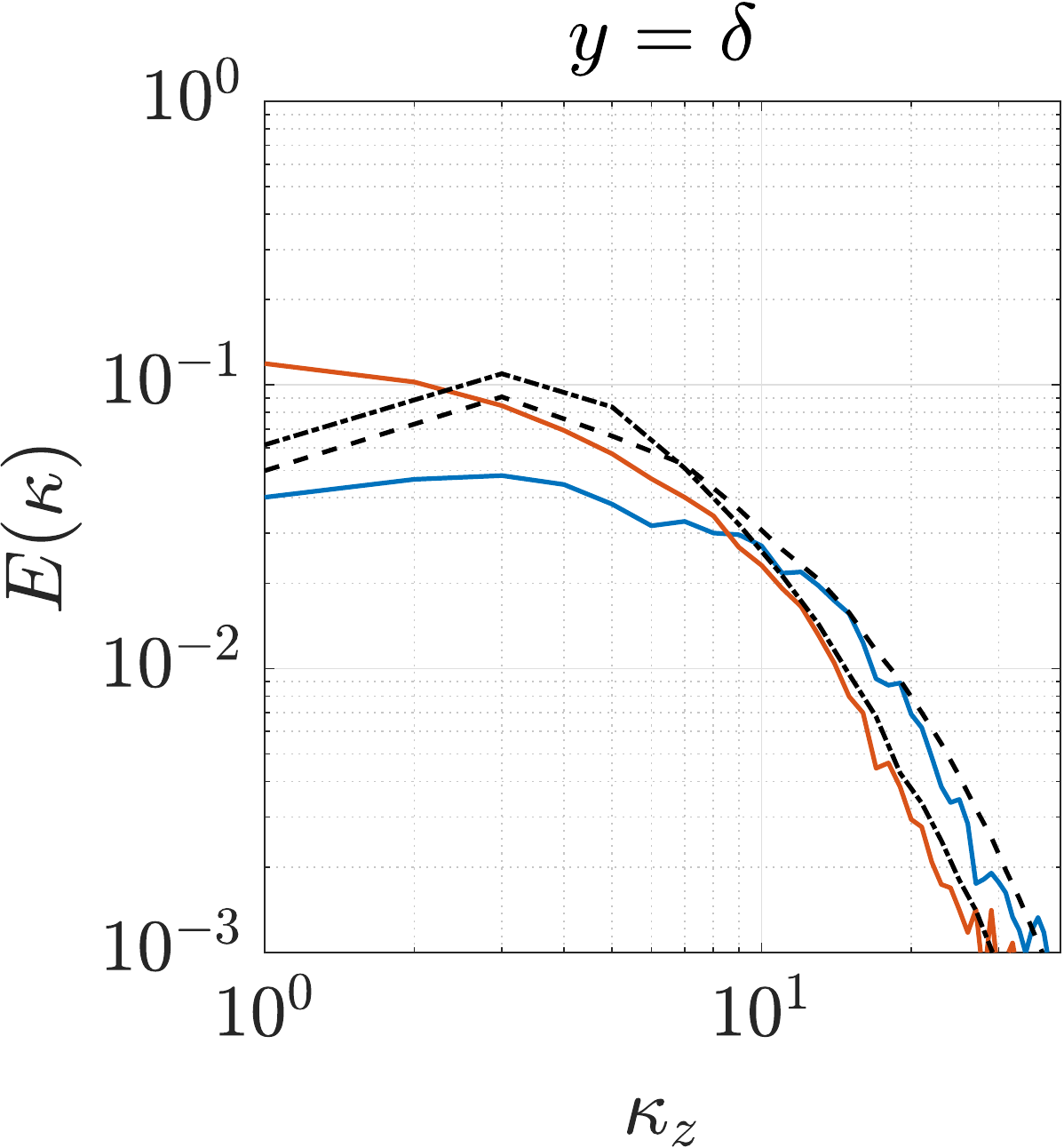}  \\
      \hspace{1.25mm} a)  \hspace{47.mm}  b)  \hspace{47.mm} c)
 \end{center}
  \caption{Turbulent kinetic energy spectra, captured at the inlet and at three different heights from the wall , compared with the DNS data collected by Moser et al. \cite{Moser}.\ blue) transverse;\ red) longitudinal;\ -\ -\ -) DNS transverse;\ -.-.-) DNS longitudinal.}
  \label{channel_E_inlet}
\end{figure}

The development of the friction coefficient at the walls is presented in Figure \ref{channel_Cf}. Results from the THV SEM are compared with data reported by Jarrin et al. \cite{Jarrin_1} using the original SEM (OSEM), Poletto et al. \cite{Poletto_2} using a divergence-free SEM (DFSEM), and Skillen et al. \cite{Skillen} using a SEM with improved eddy positioning and amplitude calculation (SSEM). The THV SEM shows the same spike in friction coefficient as the DFSEM and the SSEM immediately downstream of the inlet as the synthetic fluctuations begin to interact with the wall. $C_F$ then quickly settles to within $\pm2\%$ of the fully-developed value by $2\delta$ downstream. The THV SEM shows a significant reduction in the development distance over the DFSEM and a similar development distance with the SSEM.

\begin{figure}
 \begin{center}
  \mbox{
      \includegraphics[width=12.cm, clip=true]{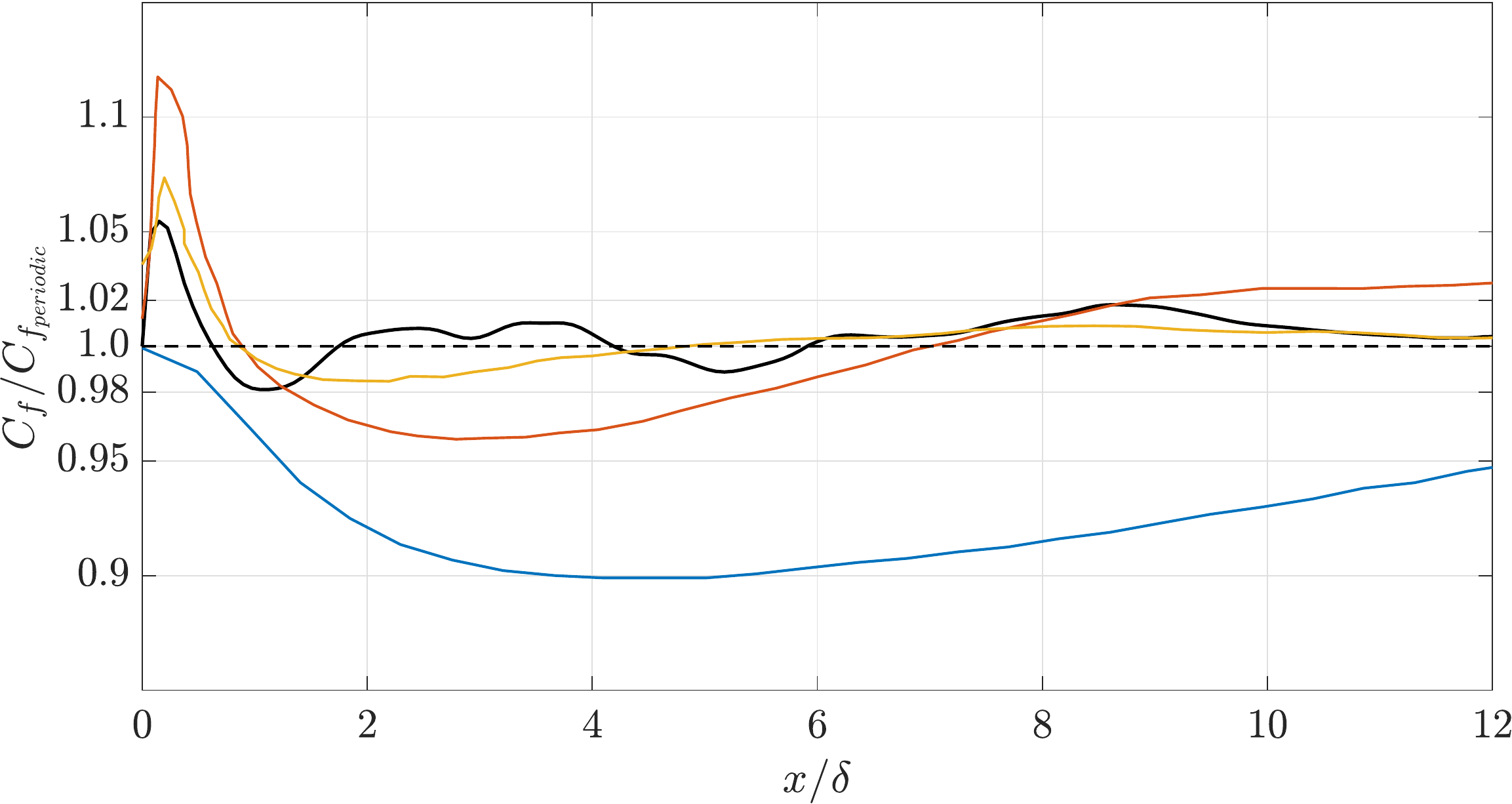}
         }  
 \end{center}
  \caption{Downstream development of the friction coefficient compared with the DNS data collected by Moser et al. \cite{Moser} and three different synthetic eddy methods:\ black) THV SEM; blue) Original SEM (OSEM) of Jarrin et al. \cite{Jarrin_1}; red) Divergence-Free SEM (DFSEM) of Poletto et al. \cite{Poletto_2}; yellow) SEM (SSEM) of Skillen et al. \cite{Skillen};\ -\ -\ -) DNS of Moser et al. \cite{Moser}.} 
  \label{channel_Cf}
\end{figure}

Vertical profiles of the turbulent kinetic energy and Reynolds shear stress at four streamwise locations are plotted in Figure \ref{channel_TKE_uv}. Both the TKE and Reynolds shear stress have recovered their given values away from the wall ($-0.5\delta<y<0.5\delta$) by one channel height downstream of the inlet ($x=2\delta$). The decrease in both quantities near the wall is because the grid in the streamwise direction is too coarse to properly resolve the smaller THV's in the fourth and fifth THV generations. Returning to Figure \ref{channel_u_vort_mag}, the larger THV's away from the wall do influence this nearer-wall region; which is why the TKE and Reynolds shear stress both decrease to values consistent with the Reynolds stresses at the center of the larger THV's. By $10\delta$ downstream of the inlet, the desired values are recovered in the near-wall region.

\begin{figure}
 \begin{center}
      \includegraphics[width=6.8cm, clip=true]{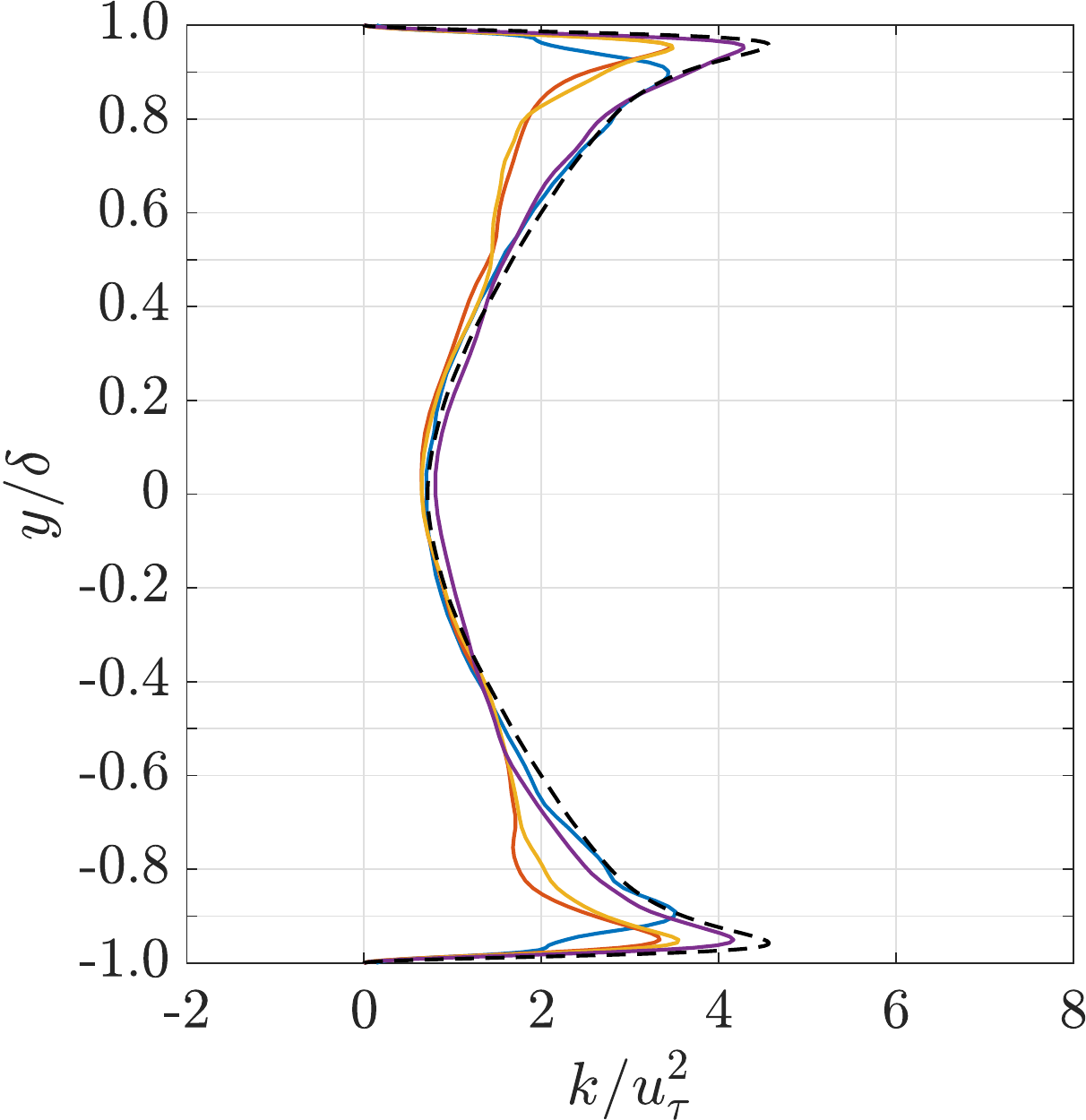}  \hspace{2mm}
      \includegraphics[width=6.8cm, clip=true]{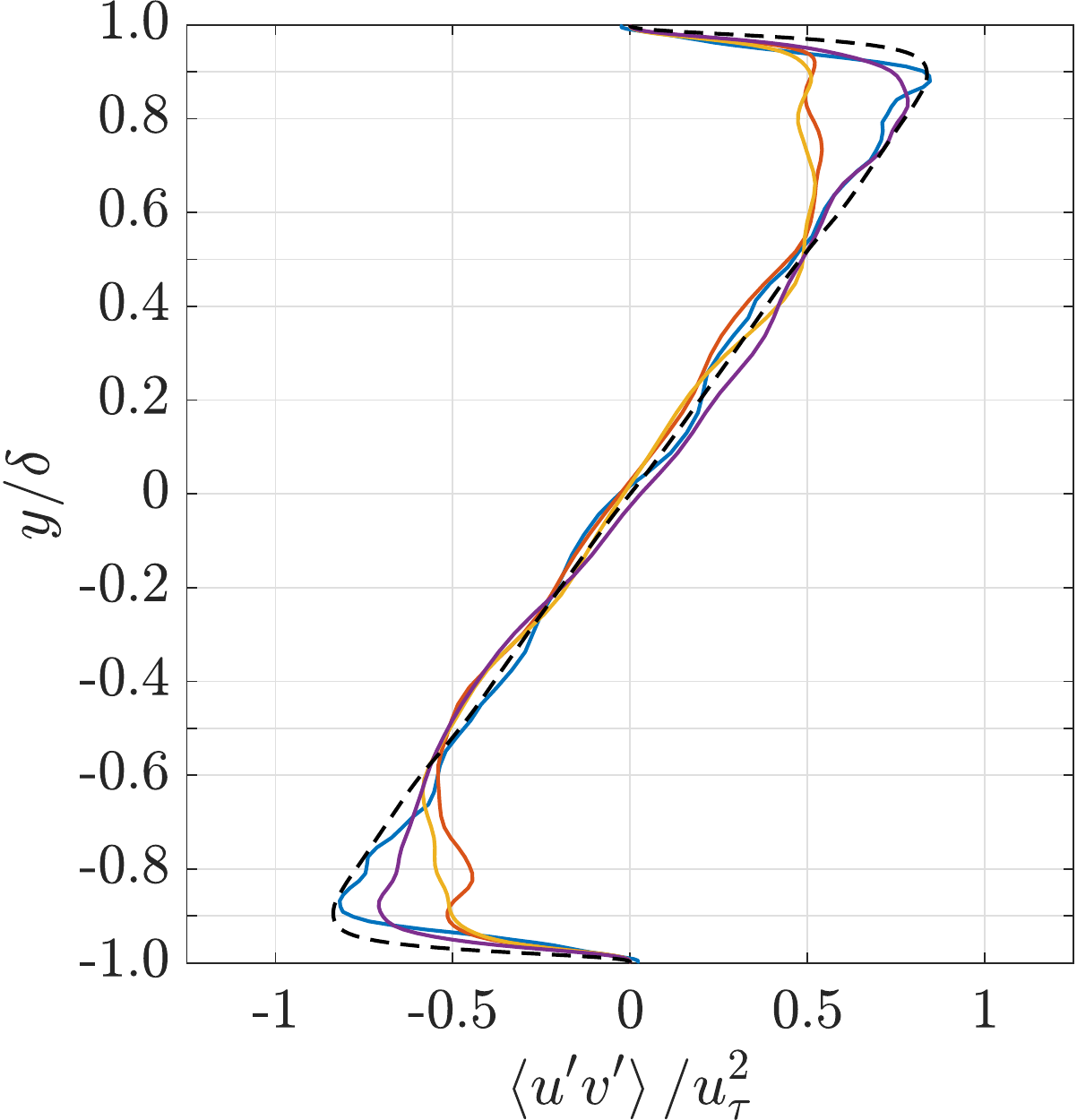}  \\
       a)  \hspace{56.25mm}  b)
 \end{center}
  \caption{Vertical profiles at four different streamwise locations compared with the DNS data collected by Moser et al. \cite{Moser} of: a) turbulent kinetic energy; b) Reynolds shear stress;\ blue) $x=0$\thinspace (inlet);\ red) $x = 2\delta$\thinspace;\ yellow) $x = 4\delta$\thinspace;\ purple) $x = 10\delta$\thinspace;\ -\ -\ -) DNS of Moser et al. \cite{Moser}.} 
  \label{channel_TKE_uv}
\end{figure}

Two-point streamwise spatial velocity correlations at two different heights from the wall, calculated from the inlet, are presented in Figure \ref{channel_Rij_x}. The influence of stretched THV's defined in Section \ref{section_stretch} can clearly be seen in the differences between the longitudinal correlation near the wall in Figure \ref{channel_Rij_x}(a) and near the height where the stretched THV's are clustered in Figure \ref{channel_Rij_x}(b). Close to the wall, only the smaller spherical THV's are imposed. Although those small THV's give good agreement at distances less than $0.5\delta$ from the inlet, their spherical nature is unable to model the larger downstream coherence. This is in stark contrast to the longitudinal correlation $0.5\delta$ away from the wall in Figure \ref{channel_Rij_x}(b). The spatial correlation is in agreement much farther downstream. This is because the stretched THV's are being created at the inlet at this general height. The stretched THV's introduce larger correlated scales which provides a better model of the physical structures.

\begin{figure}
 \begin{center}
     \includegraphics[width=6cm, clip=true]{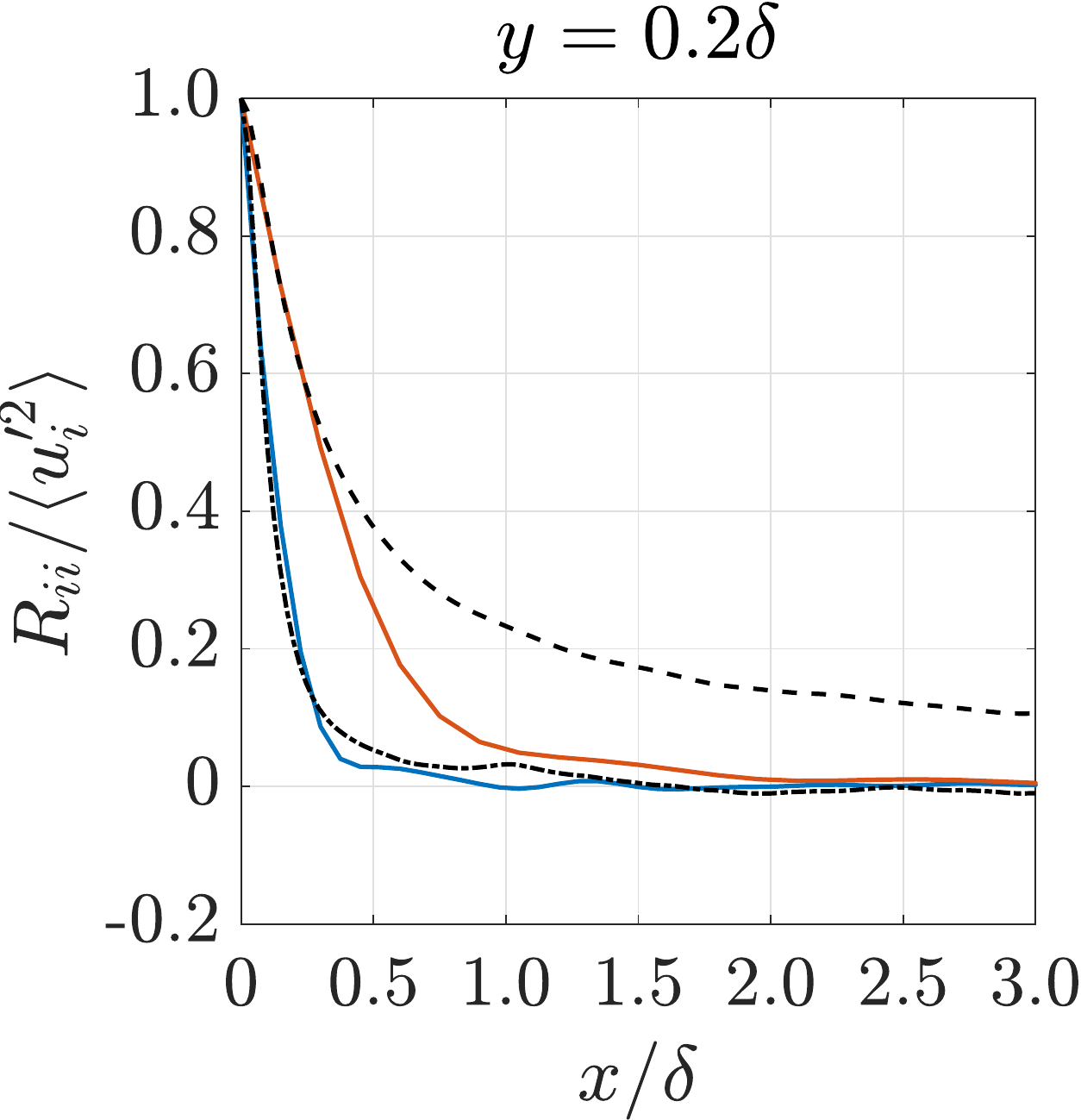}  \hspace{2mm}
     \includegraphics[width=6cm, clip=true]{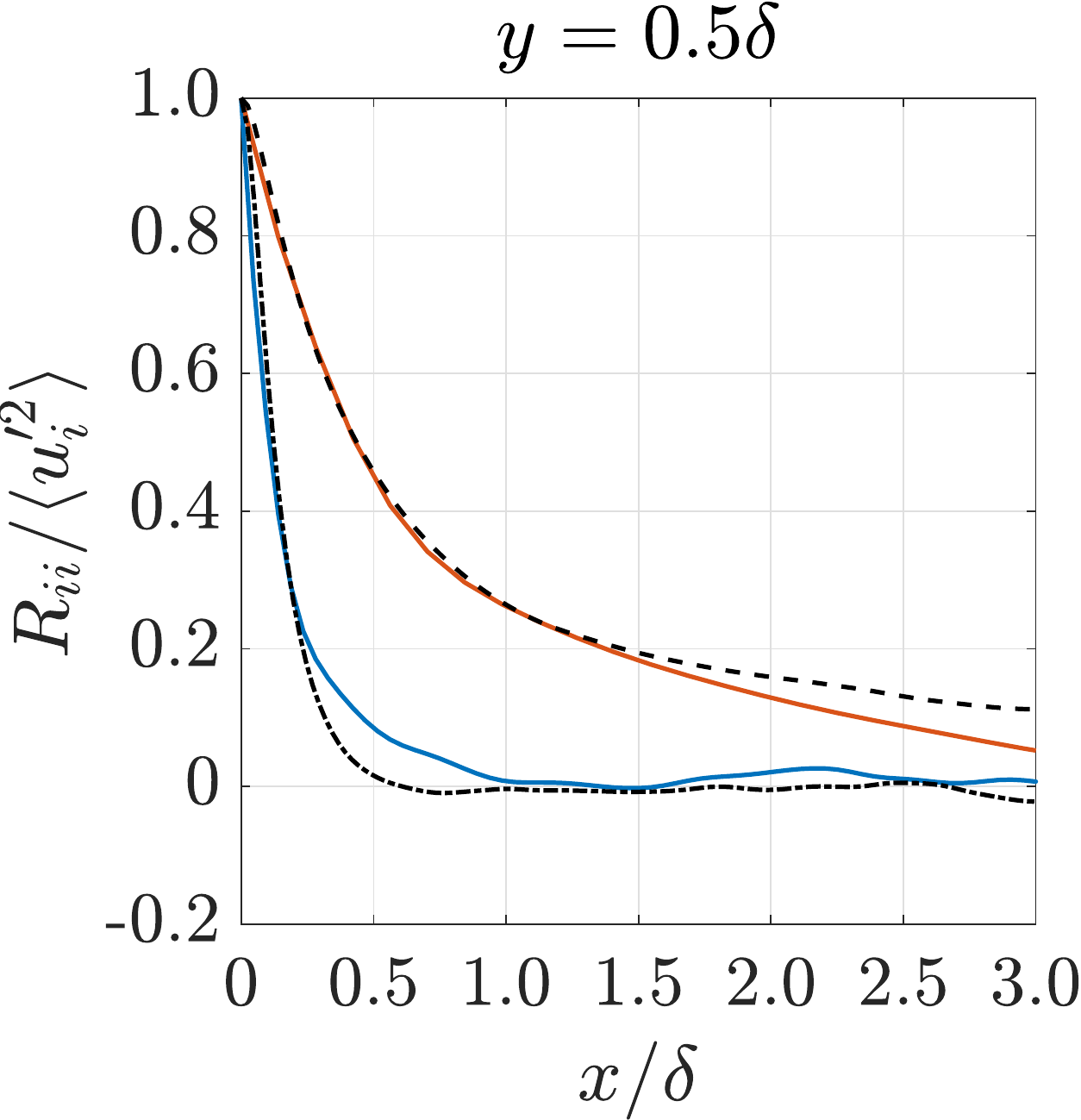}    \\  \hspace{9mm}  a)  \hspace{37.mm}  b)
 \end{center}
  \caption{Two-point streamwise spatial correlations, captured at two different heights from the wall , compared with the DNS data collected by Moser et al. \cite{Moser}. \ blue) transverse;\ red) longitudinal;\ -\ -\ -) DNS transverse;\ -.-.-) DNS longitudinal.}
  \label{channel_Rij_x}
\end{figure}

The downstream evolution of the two-point spanwise transverse spatial velocity correlation is shown in Figure \ref{channel_Rij_yz_x} and compared with DNS data from Moser et al. \cite{Moser} at two representative heights. Again, at the inlet, the THV's are well correlated at small distances, but they are unable to reproduce the larger negative correlations. As the synthetic fluctuations are transformed by the Navier-Stokes equations, the correct negative correlations quickly develop by $2\delta$ downstream of the inlet. This rapid development of the physical coherence of the synthetic fluctuations agrees with the rapid settling of the friction coefficient, the turbulent kinetic energy, and the Reynolds shear stress.

\begin{figure}
 \begin{center}
     \includegraphics[width=6cm, clip=true]{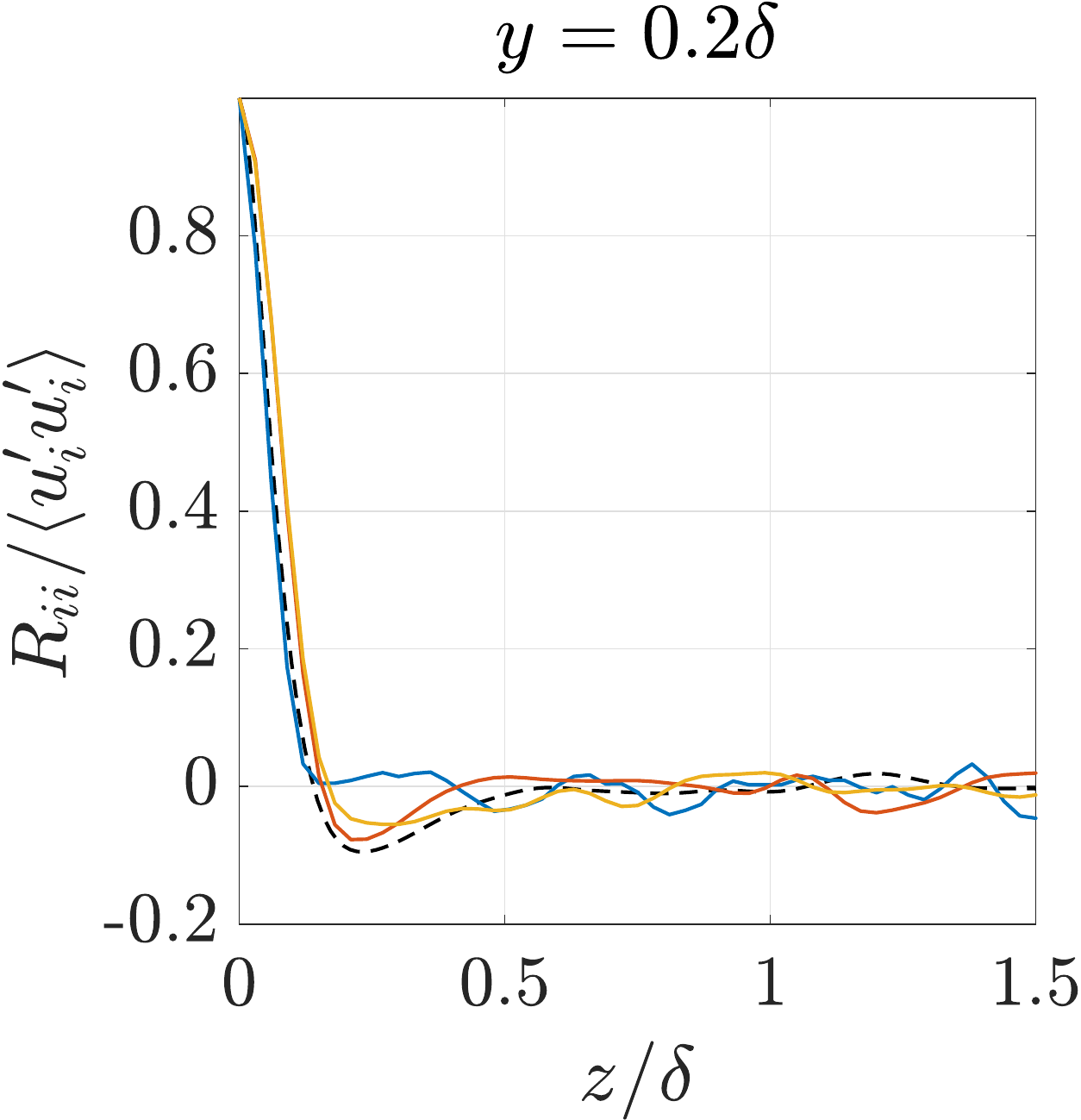}  \hspace{2mm}
     \includegraphics[width=6cm, clip=true]{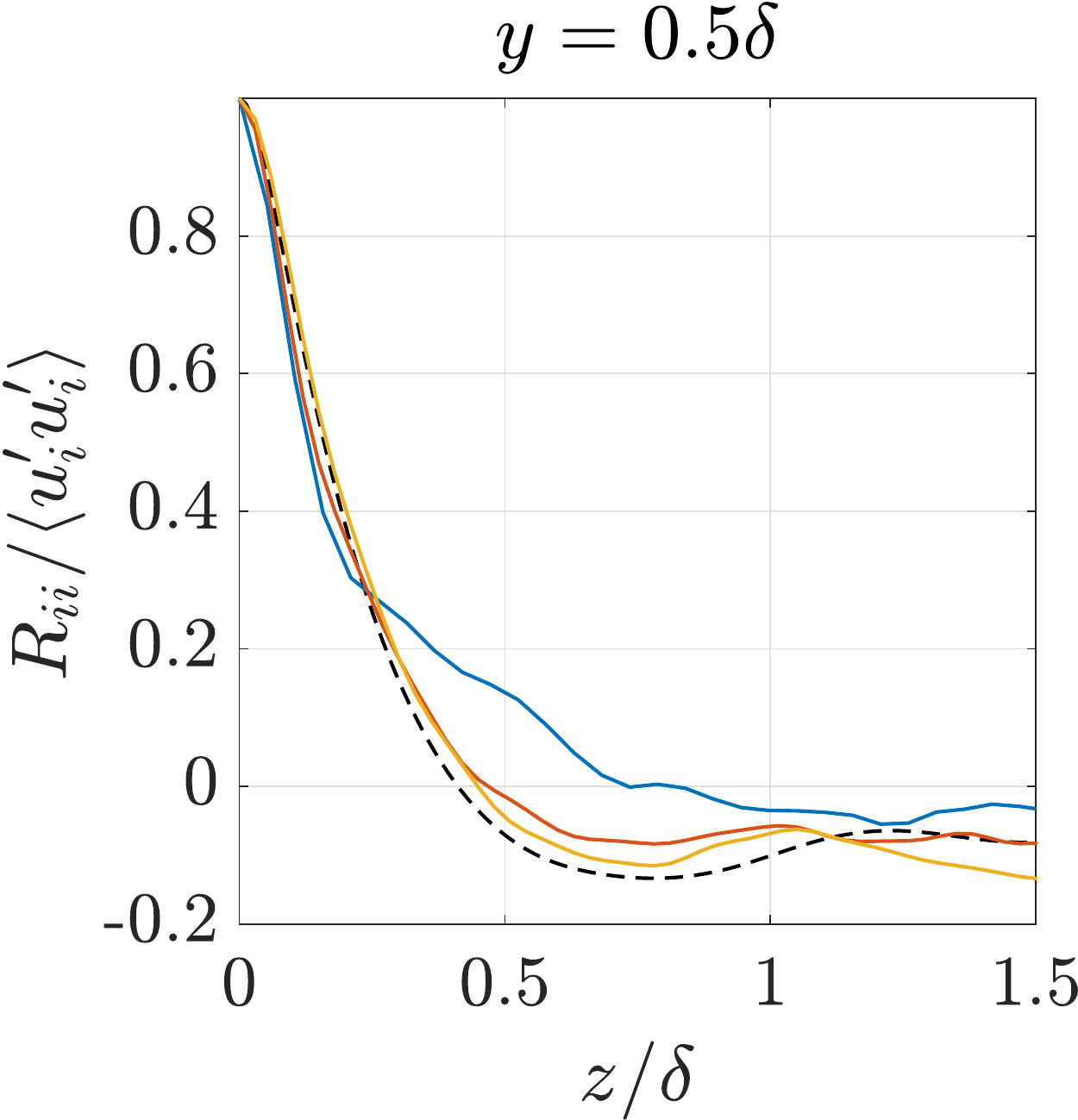}  \\
    \hspace{9mm}  a)  \hspace{37.mm}  b)  
 \end{center}
  \caption{Downstream evolution of the two-point spanwise spatial correlations, captured at two different heights from the wall, compared with the DNS data collected by Moser et al. \cite{Moser}.\ blue) $x=0$\thinspace (inlet);\ red) $x = 2\delta$\thinspace;\ yellow) $x = 4\delta$\thinspace;\ -\ -\ -) DNS transverse.}
  \label{channel_Rij_yz_x}
\end{figure}

\section{Conclusion}
The Triple Hill's Vortex was proposed as a superposition of three Hill's vortices with their axes perpendicular to each other. The proposed new synthetic eddy was applied in the framework of the synthetic eddy method in order to generate synthetic turbulent inflow velocity fields that satisfied the divergence-free condition and matched given Reynolds stress profiles. Simulation of homogeneous isotropic turbulence and turbulent channel flow were both able to reproduce given Reynolds stress tensors. The transition from artificial to realistic turbulence in the proximity to the inflow boundary was found to be small in all test cases that were considered. Excellent agreement between the turbulent kinetic energy spectrum and the two-point spatial correlations were found for the homogeneous case both at the inlet and far downstream. The spatial decay of the turbulent kinetic energy for isotropic homogeneous turbulence was shown to be in agreement with experimental data of isotropic turbulence in a wind tunnel and the synthetic fluctuations quickly developed skewness downstream of the inlet.  

For the channel flow, Reynolds stress profiles taken from DNS data were able to be reproduced away from the wall, but the distribution of Triple Hill's Vortices was not sufficient to provide good agreement very close to the walls. The recovery of the friction coefficient was shown to be as just quick as for one of the latest synthetic eddy methods. Although good agreement with the DNS data was found at the inlet for the spanwise two-point correlations, recovery of the correct correlations occurred quickly downstream.

For the future, the THV SEM needs to be expanded to allow for the use of a wider range of arbitrary turbulent statistics as inputs. The near-wall stretching and the uniform division of the total TKE across the THV generations both showed that rudimentary indirect control over the spatial correlations and energy spectra can be exerted. Through a systematic approach, general relationships might be able to be formulated to connect arbitrary energy spectra to how the target TKE for each THV generation is allocated and to connect arbitrary spatial correlations to the stretching of the THV's. The intrinsic coherence of a synthetic eddy, where all of the resolved scales smaller than the size of the eddy are contained within the eddy, introduces an additional layer of complexity that could hinder the formulation of the general relationships described above. To introduce non-zero higher order statistics (for example, skewness or flatness) into the final synthetic turbulence field, the streamfunction for the THV could be distorted such that the THV is no longer symmetric while preserving the divergence-free condition.


%
%

\section*{Funding Sources}

Effort sponsored by the Engineering Research \& Development Center under Cooperative Agreement number W912HZ-17-C-0021.  The views and conclusions contained herein are those of the authors and should not be interpreted as necessarily representing the official policies or endorsements, either expressed or implied, of the Engineering Research \& Development Center.

\section*{Disclaimer}
Reference herein to any specific commercial company, product, process, or service by trade name, trademark, manufacturer, or otherwise, does not necessarily constitute or imply its endorsement, recommendation, or favoring by the United States Government or the Department of the Army (DoA). The opinions of the authors expressed herein do not necessarily state or reflect those of the United States Government or the DoA, and shall not be used for advertising or product endorsement purposes.

\appendix 
\section{Triple Hill's Vortex Synthetic Eddy Method Outline}
This appendix serves as an outline for a general implementation of the THV SEM at an inflow plane with static Reynolds stresses. 

\noindent Before time marching, 
\begin{enumerate}
	\item Set the radius ranges and populations of the THV generations. Also, set which generations are stretched or clustered is applicable. 
	\item Set the target Reynolds stress profiles	
	\item Create all of the THV's. For each THV:
	\begin{enumerate}
		\item Assign a random radius within the bounds of the generation size
		\item Assign a random center location within the bounds of the generation clustering location, if applicable
		\item Calculate the three amplitudes
		\begin{enumerate}
			\item Calculate the Reynolds stress contribution of the THV's already present on the inflow plane at the location of the new THV
			\item Modify the target Reynolds stresses at the center of the new THV with the calculated contribution Reynolds stresses
			\item Calculate the principal Reynolds stresses and eigenvectors 
			\item If any of the principal Reynolds stresses are negative, assign a new random center location and recalculate the amplitudes at the new location. Negative principal Reynolds stresses indicate that the target Reynolds stresses are already satisfied.	 
		\end{enumerate}
		\item Multiply each of the principal Reynolds stresses by an independent random number
	\end{enumerate}
\end{enumerate}

\noindent During time marching within the calculation of boundary conditions,
\begin{enumerate}
	\item Check if current THV's need to be released (if they are over two radii downstream of the inflow plane, for example)
	\item Create new THV's to replace released THV's using the above creation steps
	\item Calculate the velocity contribution from each eddy at each grid point on the inflow plane (contributions of THV's a certain distance away, for example two radii, from a grid point are assumed to be zero)
	\item Add the fluctuating velocity to a given mean flow and impose the combination at the inflow
\end{enumerate}



\end{document}